\documentclass[aip,jcp,amsmath,amssymb,reprint,numeric]{revtex4-1}

\title{pinned system}
\usepackage{graphicx}
\usepackage{subfigure}
\usepackage{dcolumn}
\usepackage{bm}
\usepackage{mathrsfs}
\usepackage[dvipsnames]{xcolor}
\usepackage[normalem]{ulem}
\usepackage{caption}
\usepackage{comment} 

\begin{document}

\title{Exploring the soft pinning effect in the dynamics and the structure dynamics correlation in multicomponent supercooled liquids} 
\author{Ehtesham Anwar}
\thanks{\textit{Ehtesham Anwar and Palak Patel contributed equally to this work}}
\address{\textit{Polymer Science and Engineering Division, CSIR-National Chemical Laboratory, Pune-411008, India}}
\author{Palak Patel}
\thanks{\textit{Ehtesham Anwar and Palak Patel contributed equally to this work}}
\address{\textit{Polymer Science and Engineering Division, CSIR-National Chemical Laboratory, Pune-411008, India}}
\affiliation{\textit{Academy of Scientific and Innovative Research (AcSIR), Ghaziabad 201002, India}}
\author{Mohit Sharma}
\address{\textit{Polymer Science and Engineering Division, CSIR-National Chemical Laboratory, Pune-411008, India}}
\affiliation{\textit{Academy of Scientific and Innovative Research (AcSIR), Ghaziabad 201002, India}}
\author{Sarika Maitra Bhattacharyya}
\email{mb.sarika@ncl.res.in}
\address{\textit{Polymer Science and Engineering Division, CSIR-National Chemical Laboratory, Pune-411008, India}}
\affiliation{\textit{Academy of Scientific and Innovative Research (AcSIR), Ghaziabad 201002, India}}

\begin{abstract}
We study multicomponent liquids by increasing the mass of $15\%$ of the particles in a binary Kob-Andersen model. We find that the heavy particles have dual effects on the lighter particles. At higher temperatures, there is a significant decoupling of the dynamics between heavier and lighter particles, with the former resembling a pinned particle to the latter. The dynamics of the lighter particles slow down due to the excluded volume around the nearly immobile heavier particles. Conversely, at lower temperatures, there is a coupling between the dynamics of the heavier and lighter particles. The heavier particles' mass slows down the dynamics of both types of particles. This makes the soft pinning effect of the heavy particles questionable in this regime. We demonstrate that as the mass of the heavy particles increases, the coupling of the dynamics between the lighter and heavier particles weakens. Consequently, the heavier the mass of the heavy particles, the more effectively they act as soft pinning centres in both high and low-temperature regimes. A key finding is that akin to the pinned system, the self and collective dynamics of the lighter particles decouple from each other as the mass of the heavy particles has a more pronounced impact on the latter. We analyze the structure dynamics correlation by considering the system under the binary and modified quaternary framework, the latter describing the pinned system. Our findings indicate that whenever the heavy mass particles function as soft pinning centres, the modified quaternary framework predicts a higher correlation.
\end{abstract}

\maketitle
\section{Introduction}
The behavior of a liquid that has been cooled below its crystallization temperature and entered the supercooled regime is significantly different from that of a liquid at high temperatures \cite{Theorectical_perspective_berthier_2011,stillinger,ANGELL,sastry_nature_1998,Debenedetti_book}. In this supercooled liquid regime, the viscosity and the dynamics grow by several orders in magnitude before the system undergoes a glass transition\cite{ANGELL,stillinger,Debenedetti_book}. The origin of this slowing down of the dynamics has been a topic of intense research and is not completely understood. One school of thought is that the slowing down of the dynamics is due to the influence of the underlying potential energy landscape, and the divergence of the dynamics is connected to the vanishing of the configurational entropy at the Kauzmann temperature $T_{K}$ \cite{andrea_supercooled_liq_2009,Kauzmann,kirk_woly1,kirk_woly2,kirk_thirumalai,krick_thirumalai_wolynes,xia_woly_pnas}. In typical glass-forming liquids, the Kauzmann temperature, also referred to as the ideal glass transition temperature, is estimated by extrapolating data from a slightly higher temperature range where the system can be equilibrated. However, a novel technique of randomly pinning some particles in the system made it possible to access the ideal glass transition temperature $T_{K}$\cite{walter_original_pinning,smarajit_chandan_dasgupta_original_pinning,Biroli_phase_diagram,Cytoskeletal_Pinning_2015,effect_of_random_pinning,palak_ujjwal_JCP}. 

There has been a large body of literature dedicated to understanding how the dynamics and thermodynamics change with this randomly pinning of particles in a system\cite{walter_original_pinning,smarajit_chandan_dasgupta_original_pinning,palak_ujjwal_JCP,effect_of_random_pinning,parisi_jamming_pinned_system,walter_anh,paddy}. Pinning a few particles slows down the dynamics of the mobile particles and the configurational entropy of the system is found to disappear at temperatures where the particles are still mobile \cite{smarajit_chandan_dasgupta_original_pinning,reply_by_chandan_dasgupata}. It was argued that the signature of the thermodynamic transition is in the collective part of the dynamics \cite{reply_by_kob,walter_anh}. 
However, since the collective dynamics never decays completely thus, we cannot obtain a timescale of decay, and the connection of the slowing down of the dynamics with the decrease in the configuration entropy is still a debatable topic\cite{walter_original_pinning,smarajit_chandan_dasgupta_original_pinning,palak_ujjwal_JCP,reply_by_chandan_dasgupata,reply_by_kob,walter_anh}.

The dynamics of the supercooled liquid becomes even more complex for multicomponent systems\cite{smarajit_soft_pinning_pnas_2024,copling_plasma_2017,soft_pinning_of_liquid_2015,Cytoskeletal_Pinning_2015,interleaflet_coupling_2011,saurish,sayantan_fickian_2017,paddy}, specially if the components have a significant separation of timescales in their dynamics.
This kind of scenario is quite often encountered in biological systems where there are proteins and lipids \cite{copling_plasma_2017,Cytoskeletal_Pinning_2015,G_ayappa_2020,G_ayappa_2022}
and also in solutions with bigger solute particles\cite{smarajit_soft_pinning_pnas_2024}.
What happens in these complex systems can be understood better by creating some simpler model multicomponent systems where the components have differences in their timescales. One such model system was created by adding some heavy particles in the system \cite{sayantan_fickian_2017}. 
 For the lighter particles that move faster, the heavy, slower particles appear to be pinned particles over a certain timescale and act as soft pinning centres. Many of the properties of the pinned system should be observed in these soft pinned systems. 
 However, it was also shown that in such a system at low temperatures, the dynamics of the heavier and lighter particles couple\cite{saurish}. This coupling should alter the soft-pinning effect of the heavier particles. Thus, at higher temperatures where the dynamics is decoupled, the heavy particles may act as pinned particles \cite{sayantan_fickian_2017}, and at lower temperatures, the mass of the heavy particles will slow down the dynamics of both heavy and light particles \cite{saurish}. This will give rise to a complex landscape of dynamics.

Moreover, at temperatures where the dynamics of the lighter and heavier particles are decoupled, this system can act as a model system for the system where particles are randomly pinned.
Unlike the randomly pinned system, which loses its translational invariance, the system with heavy mass particles will retain the translational invariance. 
Thus, the insight gained from the study of the system with heavy mass particles can be used to understand some of the puzzling results obtained in the pinned system \cite{reply_by_chandan_dasgupata,reply_by_kob}.

In this article, we present a study of multicomponent systems where randomly $15\%$ of the particles are chosen to have heavy mass while maintaining the same size as the particles in the original system, the Kob-Andersen model \cite{Kob_Andersen_1995}. We study the effect of the mass of the heavy particles on the dynamics of the lighter particles by varying the mass of the former. We show that the heavy mass particles have two different effects on the lighter particles. At high temperatures where there is a significant separation in the timescale of the dynamics of the heavier and lighter particles, the former acts as a soft pinning centre, and the dynamics of the latter slows down due to the excluded volume around the heavy particles. As suggested earlier, at lower temperatures, the dynamics of both kinds of particles couple \cite{saurish}. However, what we find is that with an increase in the mass of the heavy particles, this coupling becomes weaker. Thus, beyond certain values of the heavy mass particles, they act as soft pinning centres both at high and low temperatures. 
One of the most interesting results of this study is that we show that the presence of the heavy mass particles slows down the collective dynamics of the lighter particles more than the self dynamics. This result is similar to that suggested for the randomly pinned system \cite{reply_by_kob}. However, since the collective dynamics in the pinned system does not show a complete decay, there was a debate about whether the collective dynamics is genuinely slower than the self dynamics \cite{reply_by_chandan_dasgupata}. 

We then study the effect of the heavy mass particles on the structure dynamics correlation of the lighter particles. As we show here, increasing the mass of the particles does not affect the structure of the liquid. Thus, we can treat the system as a binary system. However, recently, we have shown that for a system where particles are pinned, even when there is no change in the structure, there is a change in the structural properties \cite{palak_JCP_2024}. In the calculation of the structural properties like the pair excess entropy and the mean-field caging potential \cite{manoj_PRL_2021,mohit_paper1}, the system needs to be treated under the modified quaternary framework \cite{palak_JCP_2024}. Here, we show that even for a soft pinned system, the structure dynamics correlation has a higher value when the system is treated like a modified quaternary system. Thus, we suggest that the structure dynamics correlation can be used as a tool to understand the pinning effect of a soft pinned system.

The rest of the paper is structured as follows: The simulation details are in section \ref{simulation_details}. Section \ref{dynamics} presents the analysis of the dynamics of the lighter particles in the system with some heavy mass particles. In section \ref{structure-dynamics correlation}, we examine the structure-dynamics correlation at the microscopic level. In section \ref{Ternary system and its soft pinning effect}, we examine the ternary system and its soft pinning effect. Section \ref{conclusion} contains a brief conclusion to sum up the work. This paper contains four Appendix sections at the end. 

\section{Simulation Details}
\label{simulation_details}

In this study, we use the well-known Kob-Andersen (KA)\cite{kob-andersen} A:B=80:20 binary mixture interacting via the Lennard-Jones (LJ) potential. The shifted and truncated LJ interaction potential in the KA model is expressed as,
\begin{equation}
\small u(r_{\alpha\gamma})=
\begin{cases}
u^{(LJ)}r_{\alpha\gamma};\sigma_{\alpha\gamma},\epsilon_{\alpha\gamma})- u^{(LJ)}(r^{(c)}_{\alpha\gamma};\sigma_{\alpha\gamma},\epsilon_{\alpha\gamma}), & r\leq r^{(c)}_{\alpha\gamma}\\
 0, & r> r^{(c)}_{\alpha\gamma}
\end{cases}
\label{ka_model}
\end{equation}
\noindent
where, $u^{(LJ)}(r_{\alpha\gamma};\sigma_{\alpha\gamma},\epsilon_{\alpha\gamma})=4\epsilon_{\alpha\gamma}[(\frac{\sigma_{\alpha\gamma}}{r_{\alpha\gamma}})^{12}-(\frac{\sigma_{\alpha\gamma}}{r_{\alpha\gamma}})^{6}]$ and $r^{(c)}_{\alpha\gamma}=2.5\sigma_{\alpha\gamma}$. Where $\alpha,\gamma$ $\epsilon$ $\{A, B\}$ and $\varepsilon_{AA}$ = 1.0, $\varepsilon_{AB}$ = 1.5, $\varepsilon_{BB}$ = 0.5, 
$\sigma_{AA}$ = 1.0, $\sigma_{AB}$ = 0.80, $\sigma_{BB}$ = 0.88. Length, energy, and time scales are measured in units of $\sigma_{AA}$, $\varepsilon_{AA}$ and $\tau=\sqrt{\frac{m\sigma_{AA}^2}{\varepsilon_{AA}}}$, respectively. We set the mass of the particles $m_{A}=m_{B}=m=1$.
We perform molecular dynamics simulation [using the Large-scale Atomic/Molecular Massively Parallel Simulator (LAMMPS) package \cite{lammps}]; we use periodic boundary conditions and Nosé–Hoover thermostat with integration timestep 0.005$\tau$. The time constant for the Nosé–Hoover thermostat is taken to be 100 timesteps. The total number density $\rho = N/V = 1.2$ is fixed, where V is the system volume, and N is the total number of particles.
For the study of the dynamics, we take N=1000, and for the isoconfigurational runs, we take N=4000.
 Before any analysis, the system is equilibrated for a period longer than 100 $\tau_\alpha$, where $\tau_\alpha$ is the system's relaxation time ($\tau_\alpha$ is defined in section \ref{dynamics}).

\subsection {Mass system}
To set up the mass system where 15\% of the particles have heavy mass, we perform the following steps. First, we equilibrate the KA system. We randomly choose 15\% of particles from the KA system's equilibrium configuration at the target temperature and then increase the particle's mass (M). During this procedure, we make sure the heavy mass sub-populations of A and B particles with mass unit ratio remain at 80:20, which is the same as the standard KA system\cite{thesis_palak}. Before any analysis, we first equilibrate the heavy mass system for a period longer than 100 $\tau_\alpha$ with the integration time step $\Delta t$ = 0.005. Note that M is the mass of the heavy particle, which is always greater than or equal to 1. For a system where the mass of the heavy particles is $10^z$, we refer to the system as $Mz$ system. In this study, we vary $z$ from 0-9, where the KA model is the M0 system.
We also study the system where $15\%$ of the particles are pinned. The pinned particles are chosen randomly from an equilibrium configuration of the KA system at the target temperature\cite{walter_original_pinning,smarajit_chandan_dasgupta_original_pinning,palak_JCP_2024,palak_ujjwal_JCP}. Details are given in Reference \cite{palak_JCP_2024}.

\subsection{Ternary system}
For the ternary model, we replace some of the A and B particles with C-type particles, which are bigger in size. This is done in such a way that A:B=80:20 and $C/N=0.15$, where N remains the same as the original KA model. All particles interact via LJ potential. The parameters for the A and B particles in the ternary system are similar to the KA model, and $ \epsilon_{AC} = \epsilon_{BC} = \epsilon_{CC} = 1.0$ and $\sigma_{CC} = 1.20$, $\sigma_{AC} = 1.10$, $\sigma_{BC} = 1.04$ and $m_{A}=m_{B}=m_{C}=1$.

 In a system, if we replace particles with bigger-sized particles, then although the number density remains the same, the packing fraction increases. To study the ternary system in conditions similar to the KA model, we first calculate the pressure, $P$ of the KA model at $\rho=1.2$, and $T=0.45$, which is $P=3.1$. At this state point, the $\alpha$ relaxation time of the KA model is $\tau_{\alpha} \approx 2200$. 
 We then perform an NPT molecular dynamics simulation of the ternary system at a pressure $P=3.1$ and cool it until the dynamics of the system is similar to that of the KA model at $\rho=1.2$, and $T=0.45$. We find that at $T=0.48$, the $\alpha$ relaxation time of the ternary system is $\tau_{\alpha} \approx 2200$. We then calculate the density of the ternary system at $T=0.48$. We find that at this low temperature, the density is $\rho=1.04$. We then perform NVT molecular dynamics simulation of the ternary system at $\rho=1.04$. As expected, the number density of the ternary system is lower than the KA binary model. Similar to the mass system, the dynamics are studied for N=1000, and the isoconfigurational runs are done for N=4000. Even for the ternary system, we study an equivalent pinned system where we assume that the type ``C" particles are pinned. The pinning protocol is similar to that discussed for the mass system.

\section{Dynamics}
\label{dynamics}
To describe the dynamics of the system, we calculate the self part, $Q_{s}$, and collective part, $Q_{c}$, of the overlap function, which is a two-point time (t) correlation function of local density\cite{walter_Tg}.
The self part is defined as, 
\begin{equation}
Q_{s}(t) =\frac{1}{N} \Big \langle \sum_{i=1}^{N} \omega (|{\bf{r}}_i(t)-{\bf{r}}_i(0)|)\Big \rangle \quad 
\label{self}
\end{equation}
\noindent where the function $\omega(x)$ is 1 if $0\leq x\leq a$ and $\omega(x)=0$ otherwise. The parameter $a$ is chosen to be 0.3, a value that is slightly larger than the size of the ``cage'' determined from the height of the plateau in the mean square displacement at intermediate times \cite{Kob_Andersen_1995}. Thus, the quantity $Q_{s}(t)$ measures whether or not at time $t$ a tagged particle is still inside the cage it occupied at $t=0$.

\begin{figure*}
 \centering
 \includegraphics[width=0.35\textwidth]{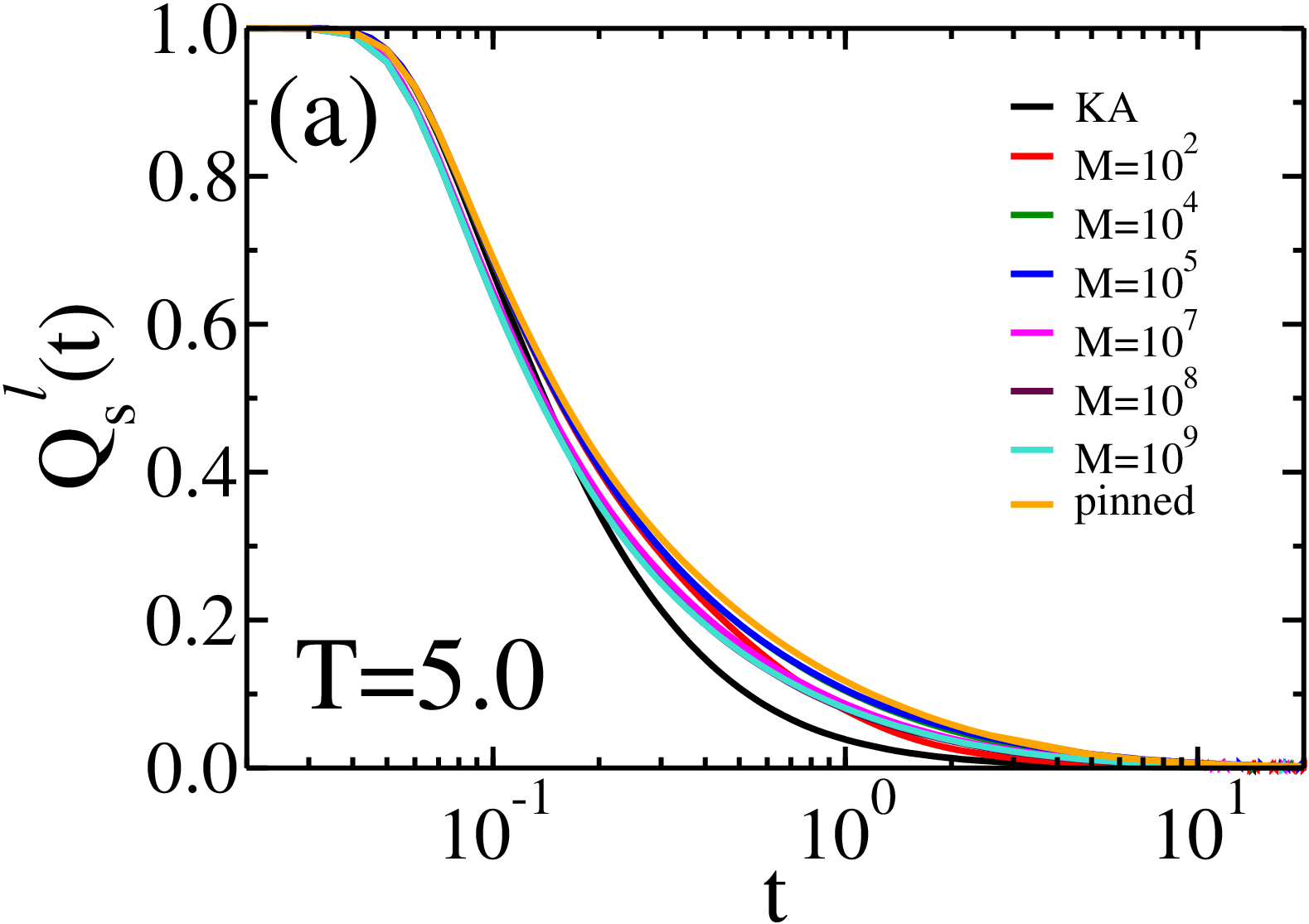}
 \hspace{0.2cm}
 \includegraphics[width=0.35\textwidth]{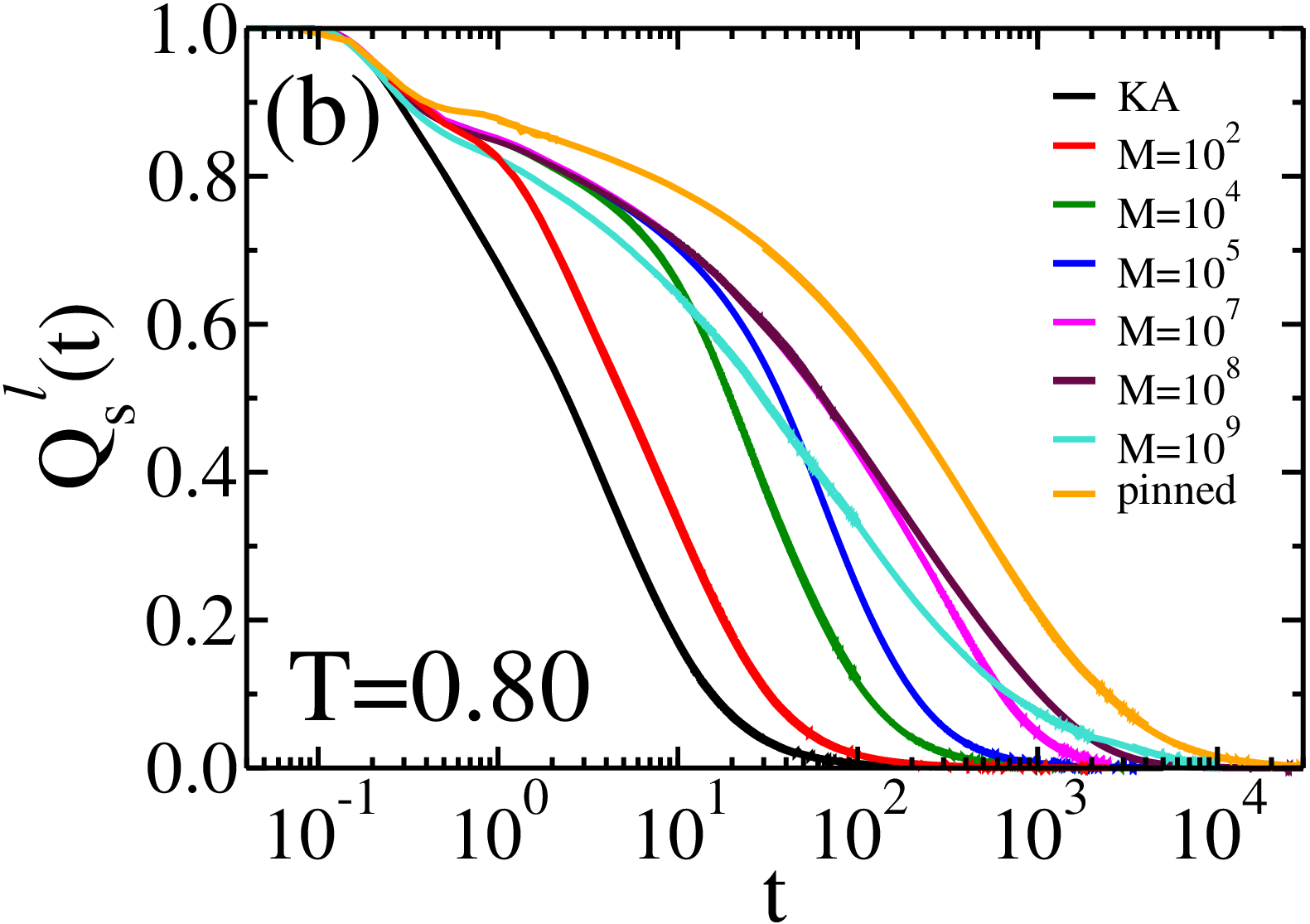}
 \\
 \vspace{0.4cm}

 \includegraphics[width=0.35\textwidth]{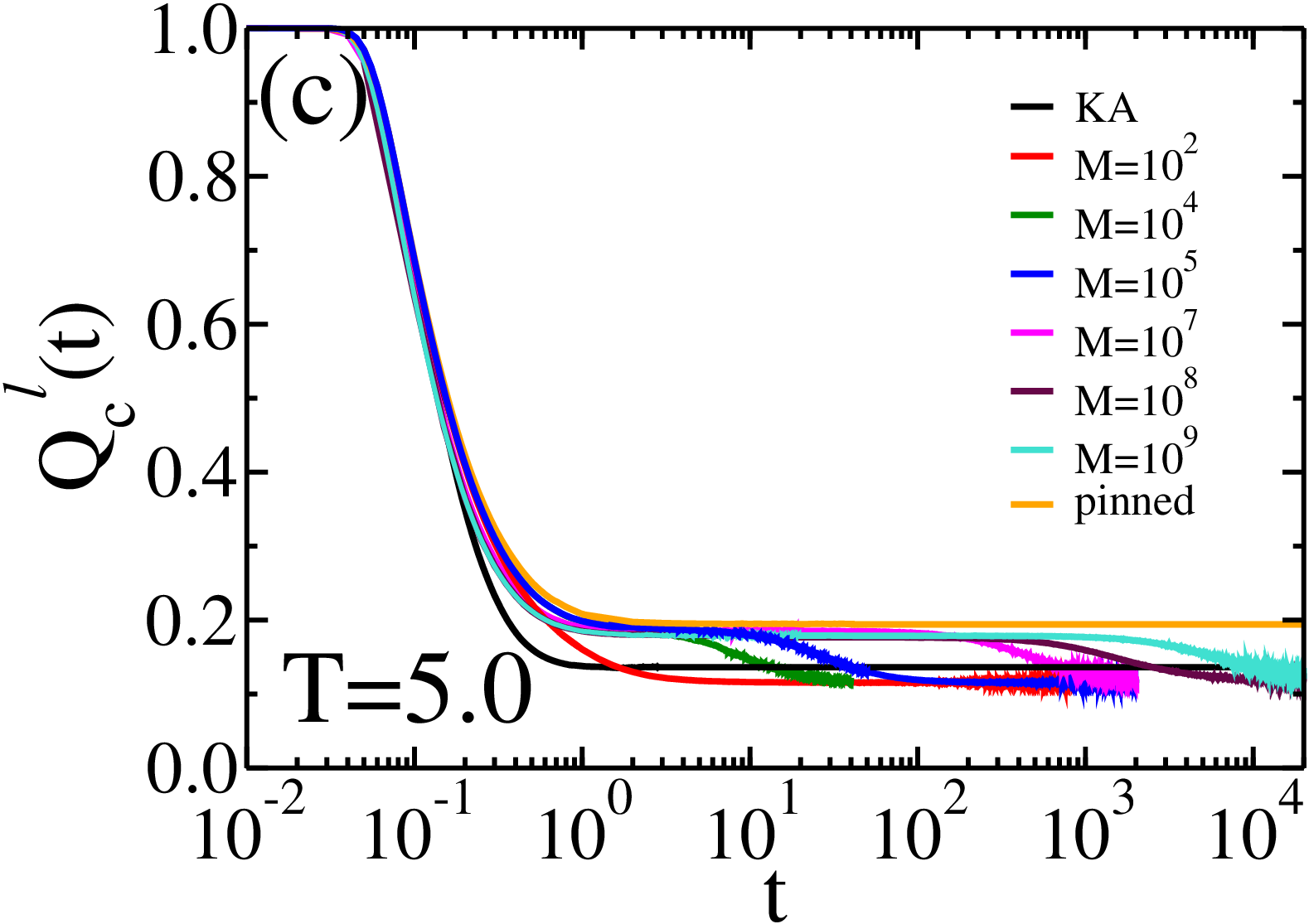}
 \hspace{0.2cm}
 \includegraphics[width=0.35\textwidth]{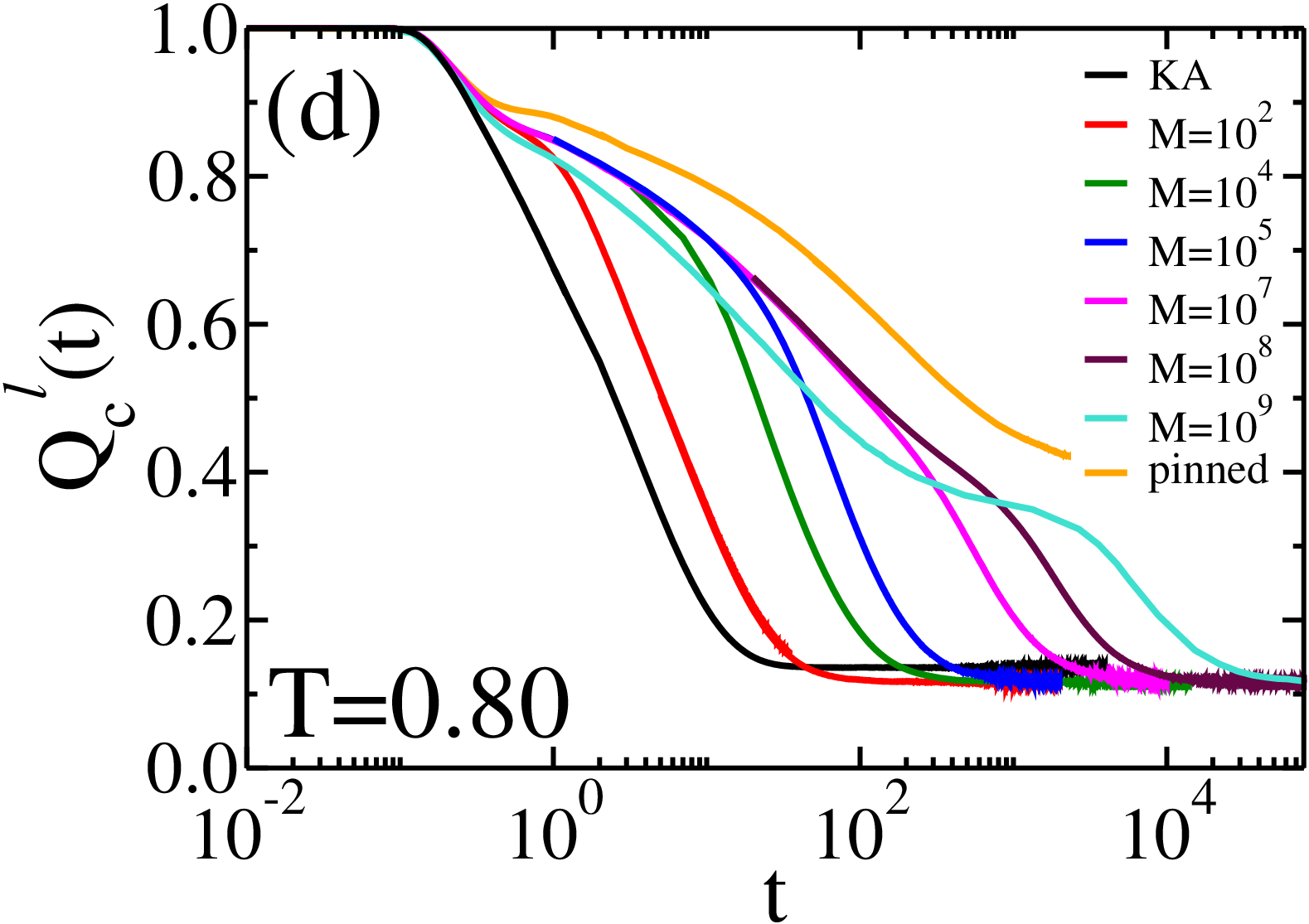}
 \\
 \vspace{0.2cm}
 \caption{The dynamic correlation functions vs. time of the lighter particles for systems with different values of mass, $M$ of the heavy particles ( KA(black) to $M=10^9$(cyan)). For comparison, we also plot the same for the dynamics of the mobile particles in the pinned system (orange) \textbf{Top Panel:} Self overlap function, $Q_{s}^{{l}}(t)$ at (\textbf{a}) T=5.0 and (\textbf{b}) T=0.80 \textbf{Bottom Panel:} Collective overlap function, $Q_{c}^{l}(t)$ at (\textbf{c}) T=5.0 and (\textbf{d}) T=0.80. Note that the collective overlap of all the particles does not decay to zero but to a value, $\lim_{t\to \infty} Q_{c}(t)=\frac{N}{V} \frac{4}{3}\pi a^{3}=0.135$ (using a=0.3) \cite{glotzer_2003,siladitya_2011}. However, the collective overlap of a fraction of the particles, here the lighter particles decays to $\lim_{t\to \infty} Q^{l}_{c}(t)=\frac{N_{l}}{V} \frac{4}{3}\pi a^{3}=0.11$.} 
 \label{light_self_collective} 
\end{figure*}

The collective overlap function is defined as follows,
\begin{equation}
Q_{c}(t) =\frac{1}{N} \Big \langle \sum_{i=1}^{N}\sum_{j=1}^{N} \omega (|{\bf{r}}_i(t)-{\bf{r}}_j(0)|) \Big \rangle \quad 
\label{collective}
\end{equation}
\noindent
 In this work, since we are interested in the pinning effect of heavy particles, we primarily investigate the dynamics of the lighter particles in the mass system. We then compare it with the mobile particles of the pinned system. We calculate the self and collective part of only the lighter particles $Q_{s}^{{l}}$ and $Q_{c}^{l}$ respectively. Thus, in Eq.\ref{self} and Eq.\ref{collective}, the calculation is over the number of lighter particles, $N=N_{l}$. We show in Fig. \ref{light_self_collective} (a) that the self overlap function for all the systems is similar at high temperatures and also similar to the pinned system. This implies that at high temperatures, the dynamics of the lighter particles are weakly affected by the heavier particles. In Fig. \ref{light_self_collective} (b), we show that the self overlap function of the lighter particles at a lower temperature (T=0.8) is different for the different systems. Thus, it appears that the mass of the heavy particles affects the dynamics of the lighter particles. 
 For comparison, we also plot the time correlation function of the mobile particles in the pinned system. We show that with the increase in the mass of heavy particles, the dynamics of lighter particles initially approach that of the mobile particles in the pinned system. However, for a very heavy mass (M9) system, we also find a speed up of the dynamics. We will discuss this point later. 
Next, we plot the collective overlap of the lighter particles in Fig. \ref{light_self_collective} (c). After an initial decay, the collective overlap shows a plateau. This plateau is very similar to that observed for the pinned system. In a short time, when the lighter particles move, but the heavier particles are yet to move, the system behaves like a pinned system, and the plateau arises due to the excluded volume at and around the pinned/heavy particle positions. Once the heavier particles start moving, there is no excluded volume, and the lighter particles can access the whole volume of the system, and the collective overlap of the lighter particles decays from the plateau. The heavier the mass, the more separation of timescale there is, and later the decay. At lower temperatures, similar to the self overlap, the dynamics of the lighter particles is strongly affected by the mass of the heavy particles (Fig. \ref{light_self_collective} (c)). Initially, with an increase in the mass of the heavy particles, the dynamics becomes slower, and for the M9 system, we find a speed up of the dynamics similar to that observed for the self overlap function.

Interestingly, unlike the self-overlap of the lighter particles, the mobility or rather the lack of it of the heavy particles appears to affect the collective overlap of the lighter particles, even at higher temperatures. This difference in the effect of heavy particles on the self and collective part of the lighter particles is an important observation and similar to that seen in the pinned system \cite{walter_original_pinning}. In the later part of this section, we will elaborate on this point.

The disappearance of the plateau in the collective overlap of the lighter particles at lower temperatures (Fig. \ref{light_self_collective}(d)) is an indication that the separation in the timescale of the dynamics of the lighter and heavier particles decreases. This happens due to the coupling of the lighter particle dynamics to the dynamics of the heavy particles. The coupling was first observed by Chakrabarty and Ni \cite{saurish}. This coupling decreases the efficiency of the heavy particles in acting as soft pinning centres. Thus, it is important to explore the dynamics of the lighter particles in the heavy mass system in greater detail and understand the regime in which the heavy mass system acts as a model system for soft pinning.

\begin{figure}[h!]
 \centering
 \includegraphics[width=0.4\textwidth]{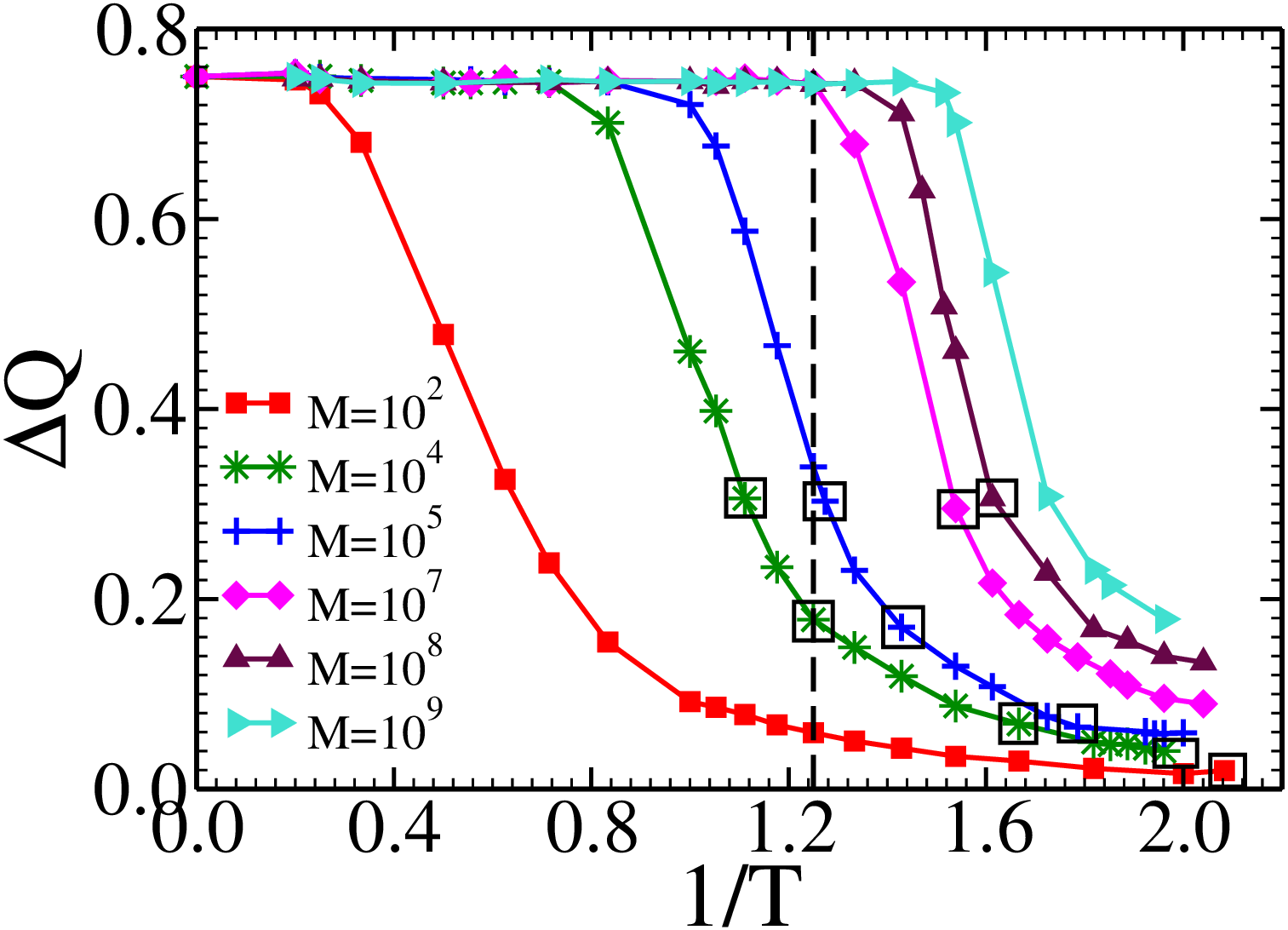}
 \caption{The temperature dependence of $\Delta Q =Q_s^{h}(t=\tau_{\alpha})-Q_s^{l}(t=\tau_{\alpha})$, where $Q_s^{h}$ and $Q_s^{l}$ are the self-overlap functions of the heavier and lighter particles, respectively. $\tau_\alpha$ is the relaxation time where the total self overlap function, $Q{_s}(t=\tau_{\alpha})=1/e$. $\Delta Q$ is plotted for different heavy mass systems ($M=10^2$(red, square) to $M=10^9$(cyan, right triangle)). The different boxes indicate the state points where the structure dynamics correlations are calculated (Fig. \ref{M2_M4}, Fig. \ref{M4_M5}, Fig. \ref{M7_M8_PIN}). The black dashed line corresponds to the onset temperature ($T_{onset} \approx 0.80$), which is the same for all the systems as suggested from the Inherent structure analysis (Shown in Appendix \ref{inherent structure}). At high temperatures, the $\Delta Q$ is almost independent of temperature. However, the $\Delta Q$ drops as the temperature is decreased, indicating a coupling between the dynamics of the heavier and lighter particles.}
 \label{deltaq}
 \end{figure}

As suggested in the earlier study \cite{saurish}, the coupling can be quantified by $\Delta Q$, which describes the difference in the dynamics of the lighter and heavier particles and is given by \cite{saurish} $\Delta Q =Q_s^{h}(t=\tau_{\alpha})-Q_s^{l}(t=\tau_{\alpha})$, where $Q_s^{h}$ is the self-overlap function of the heavy particles and $\tau_\alpha$ is the relaxation time where the total self overlap $Q{_s}(t=\tau_{\alpha})=1/e$. As shown in Fig. \ref{deltaq} at high temperatures, the $\Delta Q$ is almost independent of temperature. However, the $\Delta Q$ drops as the temperature is decreased. Similar to the earlier study\cite{saurish}, we observe that as the mass of the heavier particle increases, this drop happens at a lower temperature. We also observe that as the heavier particle's mass increases, the coupling between the dynamics of the lighter and heavier particles, even at lower temperatures, weakens. Thus, as the temperature is lowered, although the $\Delta Q$ drops from its high-temperature value, it does not decay to zero and saturates at a finite value. This saturation value increases with the mass of the heavy particles. As we will show later, due to this effect, the heavy mass particles beyond a certain mass value act as a soft pinning centre even at lower temperatures.

\begin{figure}[h!]
 \centering
 \includegraphics[width=0.45\textwidth,trim ={0 0cm 0 0.0cm},clip]{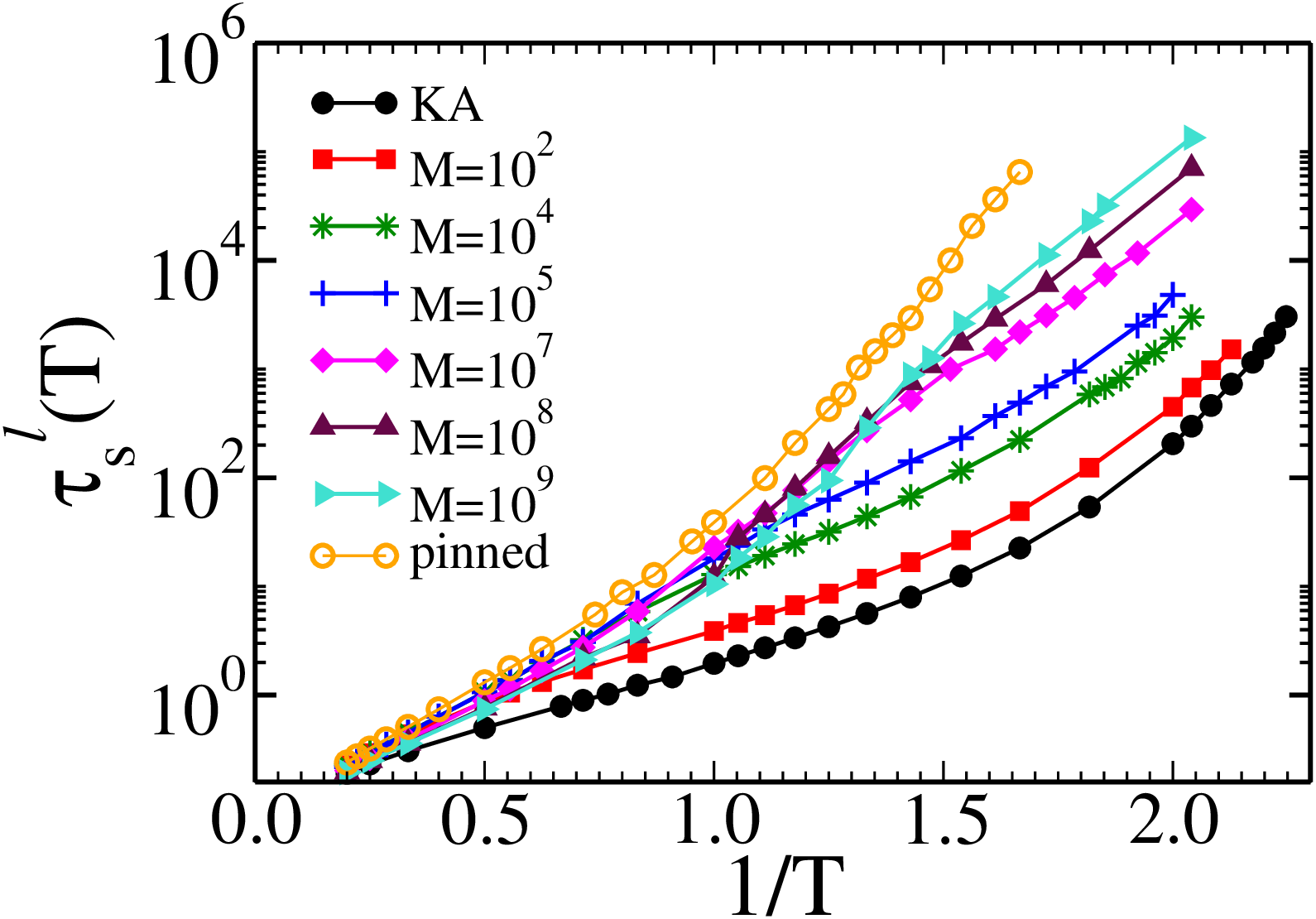}
 \caption{ The temperature dependence of the self relaxation time of the lighter mass particles, $\tau_{s}^{l}$ for the different systems (KA(black, circle) to $M=10^9$(cyan, right triangle)). For comparison we also plot the same for the mobile particles in the pinned system (orange, open circle). As the mass of heavier particles increases, the dynamics of lighter particles moves closer to that of the mobile particles in the pinned system.}
 \label{tau_light_self}
 \end{figure}

To understand the temperature dependence of the dynamics of the lighter particles and how it is affected by the mass of the heavy particles, in Fig. \ref{tau_light_self} we plot $\tau_{s}^{l}$ which is defined as $Q_{s}^{l}(t=\tau_{s}^{l})=1/e$. For comparison, we also plot the dynamics of the mobile particles in the pinned system and that of all particles in the KA model. In the temperature range presented here, the self dynamics of the lighter particles in M2, M4, M6, systems appear to grow continuously with temperature like the KA model. As we move to systems with higher mass values of the heavier particles, we find that there are two different regimes in the dynamics. This is seen clearly for M7, M8 and M9 systems. In the first part, the relaxation time of the lighter particles grows even though the dynamics of the heavier and the lighter particles are decoupled, as quantified from the $\Delta Q$ plot (Fig. \ref{deltaq}). This growth is similar to that of the pinned system and happens due to the excluded volume effect of the heavy mass particles. Eventually, in the second part, when there is a coupling in the dynamics of the lighter and heavier particles, the rate of growth of the relaxation time slows down and follows a different curve, giving rise to a kink in the $\tau_{s}^{l}$ vs. 1/T plot. With an increase in the mass of the heavier particles, this kink becomes more prominent and also shifts to lower temperatures. 
 
We find that for systems where the mass of the heavy particles is very high, like the M8 and M9 systems, in the first regime, the relaxation time marginally decreases with the increase in mass of the heavy particles. A similar observation was made in the plot of the correlation functions at $T=0.8$ (Fig. \ref{light_self_collective}). We do not completely understand the origin of this speed-up of the dynamics. 
One possibility is that for the M7 system in this first regime, there is a weak effect of the dynamics of the heavy mass particles on the lighter particles, and this effect becomes weaker as the mass of the heavy particles increases. This gives rise to a decrease in the relaxation time of the lighter particles with an increase in the mass of the heavy particles. 
Eventually, at lower temperatures where the dynamics of the heavy and light mass particles are coupled in a stronger manner, the relaxation time grows with the mass of the heavier particles.

\begin{figure}[h!]
 \centering
 \includegraphics[width=0.45\textwidth,trim ={0 0cm 0 0.0cm},clip]{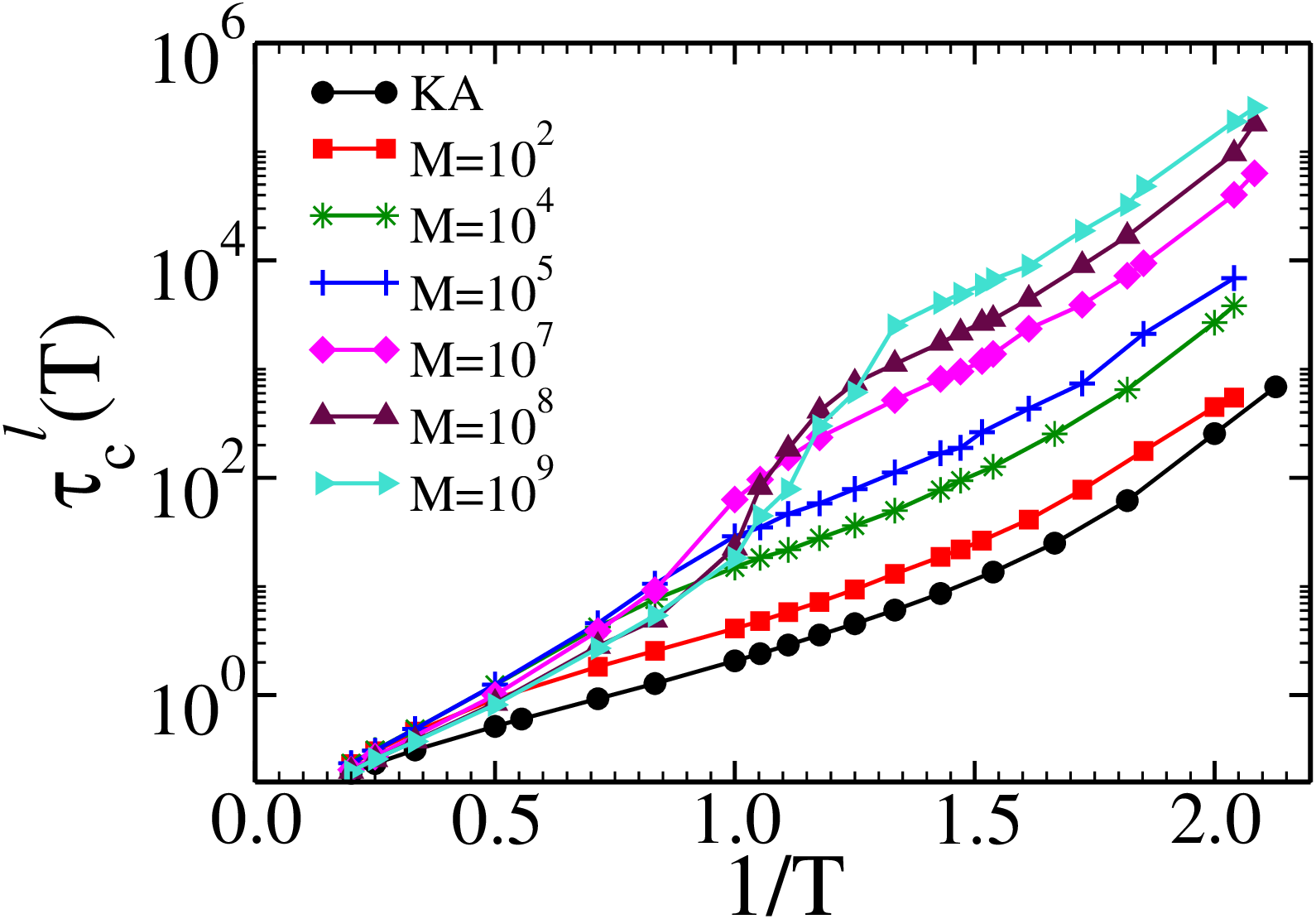}
 \caption{The temperature dependence of the collective relaxation time of the lighter mass particles, $\tau_{c}^{l}$ for the different systems (KA(black, circle) to $M=10^9$(cyan, right triangle)). Here, the plots for the systems with higher values of mass, $M$ of the heavier particles clearly show two different regions.}
 \label{tau_light_collective}
 \end{figure}

\begin{figure}[h!]
 \centering
 \includegraphics[width=0.45\textwidth,trim ={0 0cm 0 0.0cm},clip]{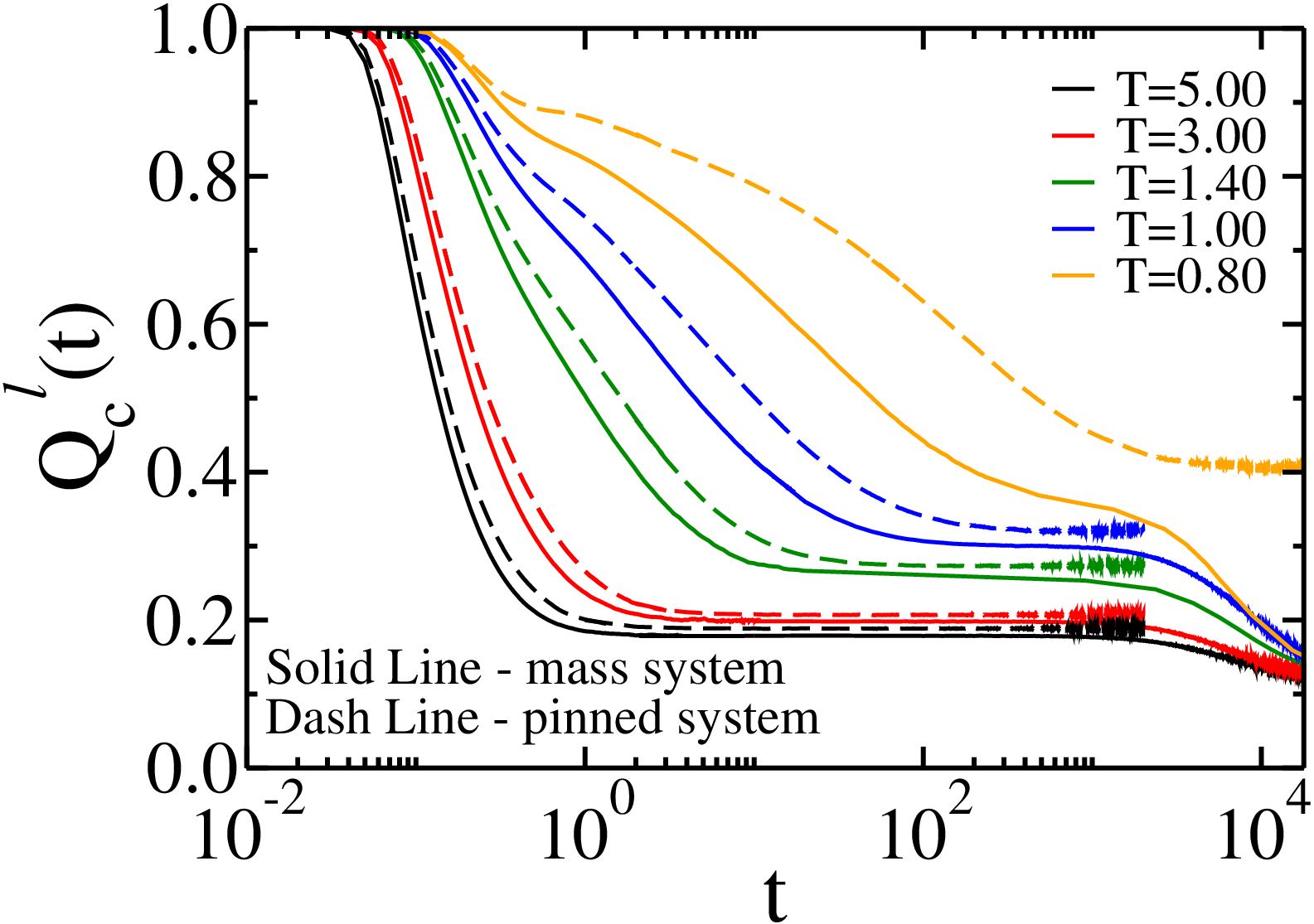}
 \caption{Collective overlap function, $Q_{c}^{l}$, (Eq. \ref{collective}) of lighter mass particles, for $M=10^{9}$ system (Solid line) and that of the mobile particles in the pinned system (Dashed line), plotted as a function of time, t at different temperatures, T.}
 \label{correlation_collective} 
 \end{figure}

In Fig. \ref{tau_light_collective}, we plot the temperature dependence of the collective relaxation time, $\tau_{c}^{l}$, of the lighter particles for the different systems, defined as $Q_{c}^{l}(t=\tau_{c}^{l})=1/e$. Similar to that observed for the self part, for systems with lower values of the mass of the heavy particles, 
the relaxation time is closer to that of the KA system, and the relaxation time grows with the increase in mass of the heavy particles. Like the self dynamics, there are two regions of growth. The kink in the $\tau_{c}^{l}$ vs. 1/T plot is even more prominent for the collective timescale. Unlike that for the self-part, we cannot compare the relaxation times of the collective part of the mass system with that of the pinned system, as we cannot calculate the timescale of the collective dynamics of the pinned system. However, in Fig. \ref{correlation_collective}, we show that the collective overlap of lighter particles in the M9 system appears quite similar to that of the mobile particles of the pinned system, especially at high temperatures. At low temperatures, due to the coupling in dynamics between the lighter and heavier particles, the time correlation function of the two systems starts showing different behavior.

\begin{figure}[h!]
 \centering
 \includegraphics[width=0.45\textwidth,trim ={0 0cm 0 0.0cm},clip]{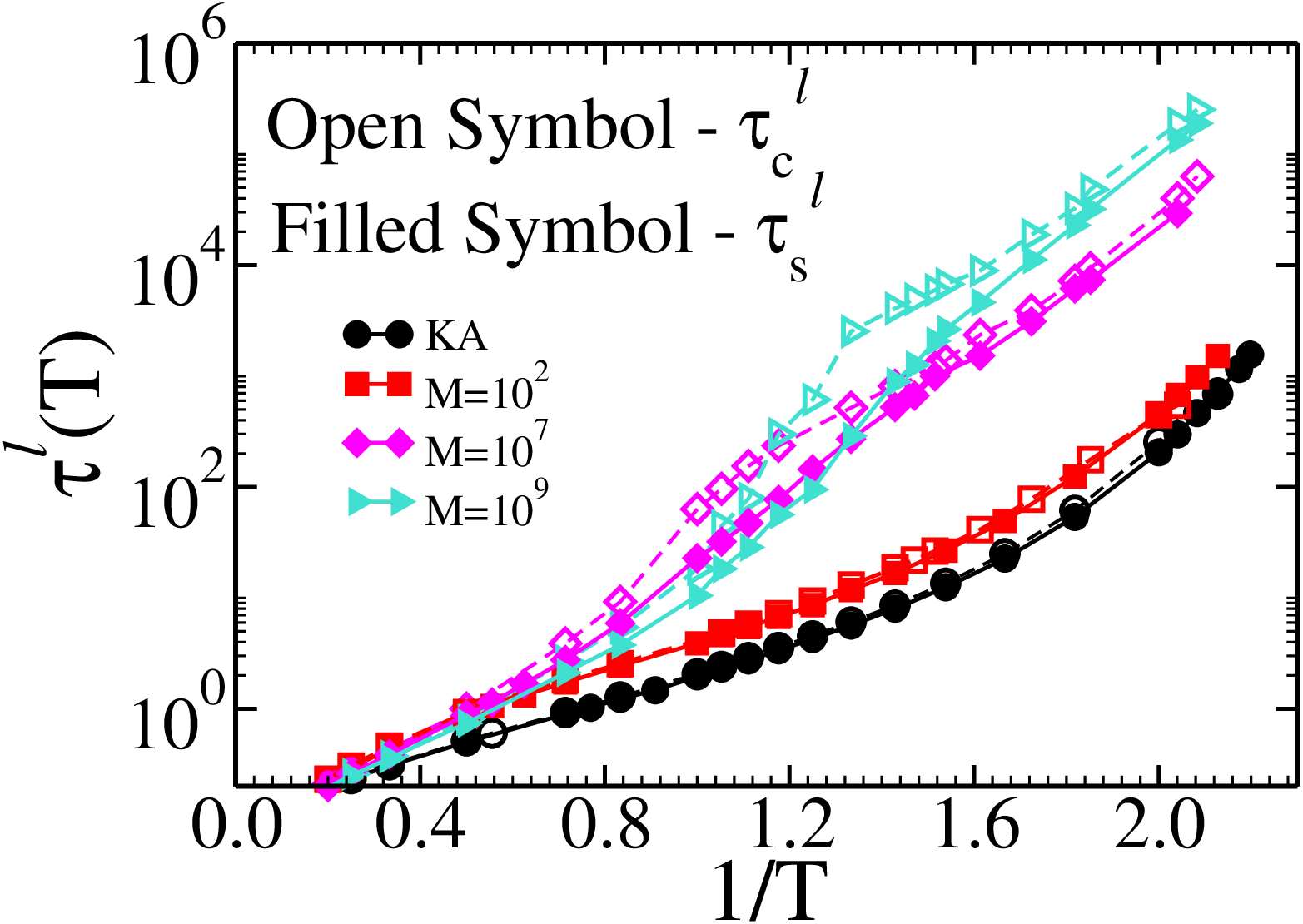}
 \caption{A comparison of the temperature dependence of the relaxation time of the lighter mass particles obtained from the self overlap function, $\tau_s^l$(filled symbol) and collective overlap function, $\tau_c^l$(open symbol) for different systems (KA(black, circle), $M=10^2$ (red, square), $M=10^7$ (magenta, diamond), and $M=10^9$(cyan, right triangle)). At intermediate temperatures, the difference between $\tau_s^l$ and $\tau_c^l$ increases with increasing the mass of heavy particles.}
 \label{tau_light_self_coll}
 \end{figure}

 In Fig. \ref{tau_light_self_coll}, we plot the temperature dependence of the self and collective relaxation timescales of the lighter particles for a few systems, namely the KA, M2, M7, and M9. We show that, for the KA and the M2 system, the self and collective timescales follow each other. However, for the M7 and M9 systems, we find that the two regions of the dynamics mentioned earlier are more pronounced for the collective dynamics compared to the self dynamics. 
 
Note that in obtaining the relaxation time, we have followed the usual protocol of choosing the relaxation time as the time when the correlation function reaches a value of 1/e.
 When a time correlation function has a continuous decay, if we obtain the relaxation time by choosing the time when the function reaches 1/e value or any other value lower than that, does change the absolute value of the relaxation time but not the other characteristics like the temperature dependence of the dynamics. However, when the dynamics of the system have a clear plateau, then the relaxation time obtained from the part of the correlation that approaches the plateau can have different behavior from the one that is obtained after the plateau. 
\begin{figure}[h!]
 \centering
 \includegraphics[width=0.45\textwidth,trim ={0 0cm 0 0.0cm},clip]{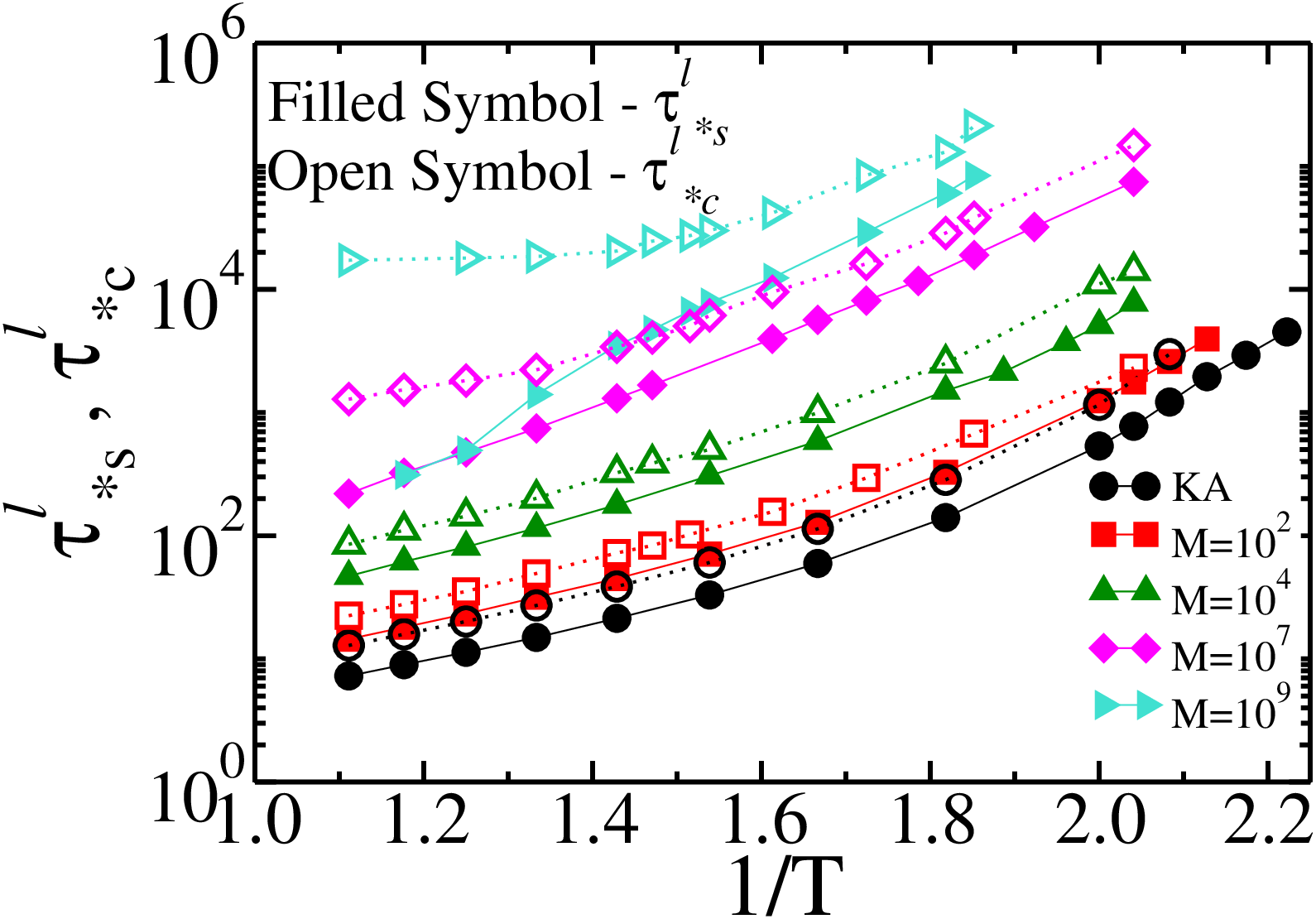}
 \caption{A comparison of the temperature dependence of the relaxation time of the lighter mass particles obtained from the self overlap function, $\tau_{*s}^{l}$ (where $Q_{s}^{l}(t=\tau_{*s}^{l})= 0.15$, filled symbol) and collective overlap function, $\tau_{*c}^{l}$ (where $Q_{c}^{l}(t=\tau_{*c}^{l})= 0.15$, open symbol) for systems with different values of mass of the heavy particles (KA (black, circle) to $M=10^9$(cyan, right triangle)) are plotted as a function of inverse of temperature, 1/T. The difference between $\tau_{*s}^{l}$ and $\tau_{*c}^{l}$ increases with increasing the mass of heavy particles.}
 \label{vft_plot_lgt_self_coll}
 \end{figure} 

 Thus, we also extract the self, $\tau_{*s}^{l}$ and collective, $\tau_{*c}^{l}$ relaxation times by choosing the time when the respective correlation functions reach an arbitrarily chosen small value of $0.15$. Thus, $Q_{s}^{l}(t=\tau_{*s}^{l})=0.15$ and $Q_{c}^{l}(t=\tau_{*c}^{l})=0.15$. In Fig. \ref{vft_plot_lgt_self_coll}, we plot some of the systems' self and collective relaxation times. We find that the two relaxation times follow each other for the KA, M2, and M4 systems. However, beyond the M7 system, there is a wide separation between the self and collective relaxation times at high temperatures. Eventually, at low temperatures, where there is a coupling between the dynamics of the lighter and heavier particles, the two relaxation times follow each other (Fig. \ref{deltaq}). With the increase in mass of the heavy particles, this separation of the self and collective relaxation times at high temperatures increases, and the temperature range where they follow each other moves to lower values. Thus, as the mass of the heavy particle increases, the self and collective dynamics of the lighter mass particles behave differently. The heavy mass particles affect the collective dynamics of the lighter particles more than their self dynamics. This effect is similar to that observed in the pinned system where the presence of the pinned particles was shown to have a more substantial effect on the collective dynamics than the self dynamics of the mobile particles \cite{walter_original_pinning,reply_by_kob}.

\begin{table}[h!]
 \centering
 \caption{Vogel–Fulcher–Tammann (VFT) temperatures $T^{s}_{VFT}$ and $T^{c}_{VFT}$ from the VFT fit of the self ( $\tau_{*s}^{l}$) and collective ($\tau_{*c}^{l}$) relaxation times, respectively, for the different systems. Here $M$ represents the mass of heavy particles in the system.}
 \begin{tabular}{|c|c|c|c}
 \hline
System & $T^s_{VFT}$ & $T^c_{VFT}$\\

\cline{1-3}
 KA & 0.3028 $\pm$ 0.002 & 0.3336 $\pm$ 0.005 & \\
\cline{1-3}
M=$10^2$ & 0.3036 $\pm$ 0.002 & 0.3005 $\pm$ 0.005 & \\
\cline{1-3}
M=$10^4$ & 0.2660 $\pm$ 0.010 & 0.2904 $\pm$ 0.011 & \\
\cline{1-3}
M=$10^5$ & 0.2211 $\pm$ 0.010 & 0.2355 $\pm$ 0.011 & \\
\cline{1-3}
M=$10^6$ & 0.1758 $\pm$ 0.011 & 0.2331 $\pm$ 0.016 & \\
\cline{1-3}
M=$10^7$ & 0.1632 $\pm$ 0.011 & 0.2653 $\pm$ 0.027 & \\
\cline{1-3}
M=$10^8$ & 0.1683 $\pm$ 0.026 & 0.3083 $\pm$ 0.032 & \\
\cline{1-3}
M=$10^9$ & 0.2523 $\pm$ 0.046 & 0.3972 $\pm$ 0.032 & \\
\cline{1-3}
 \end{tabular}
 \label{vft_self_collective}
\end{table}

Table \ref{vft_self_collective} displays the Vogel–Fulcher–Tammann (VFT) \cite{VFT,Vogel,Tammann,Fulcher} temperatures for the two relaxation times obtained by fitting the data in Fig. \ref{vft_plot_lgt_self_coll}. It is important to note that for lower values of the heavy mass, M2-M5, the dynamics at low temperatures are mainly influenced by the heavy mass (Fig. \ref{deltaq}), and the relaxation time of the lighter particles exhibits a reasonably good VFT fit. However, in systems with higher values of mass of the heavy particles (beyond M7), a single VFT fit is not feasible due to the different regions of the relaxation time. In such scenarios, we focus on fitting the low-temperature region of the data. The VFT temperatures for both self and collective dynamics, $T^{s}_{VFT}$ and $T^{c}_{VFT}$, demonstrate a non-monotonic relationship with the mass of the heavy particles. Initially, as the mass of the heavy particles increases, the VFT temperature decreases, followed by an increase. The error bars are determined by selecting various ranges for the VFT fit. Despite both $T^{s}_{VFT}$ and $T^{c}_{VFT}$ exhibiting similar trends, the difference between them grows as the mass of heavy particles increases. This difference reflects the distinct impacts that heavy particles have on the collective and self parts of the dynamics of lighter particles, with the former being more influenced by the heavy particles.

\section{Structure - Dynamics Correlation}
\label{structure-dynamics correlation}
The study of the dynamics of the lighter particles presented in the previous section shows that with an increase in the mass of the heavy particles, the mass system starts behaving more like a pinned system. However, there is also this coupling between the dynamics of the lighter and heavier particles at low temperatures. 

The mass system may act as a soft pinning model only when, for the lighter particles, the heavier particles appear to be pinned for a certain interval of time. In other words, there is a separation in timescale between the dynamics of the lighter and heavier particles. From Fig. \ref{deltaq} we find that the separation of timescale where the $\Delta Q$ is almost independent of temperature is there for a small temperature range. In this section, we use the structure dynamics correlation to understand the pinning effect of the heavy mass particles.

The correlation between the structure and dynamics can be obtained by studying the correlation between any local structural order parameter and the mobility of the particle obtained from isoconfigurational runs (IC). In simulations, the mobility of a particle at a particular time depends not only on the local structure but also on the velocity of the particle. To quantify the role of the structure on the dynamics, Harowell and coworkers designed this powerful technique, IC, whose details are given in the Appendix \ref{Isoconfiguration run} \cite{Harowell}. There are different measures of the local structure\cite{Liu15,Liu16,Liu17,Yodh19,Kohli20,Tanaka18,Tanaka18-1,Tanaka20}. In this work, we will use the local mean-field caging potential developed by some of us \cite{manoj_PRL_2021,mohit_paper1,mohit_paper2,palak_JCP_2024,palak_polydisperse_softness}. Below, we state a few important steps in the derivation of the mean-field caging potential.

The time evolution of the density under mean-field approximation can be written in terms of a Smoluchowski equation in an effective mean-field caging potential, which is obtained from the Ramakrishnan–Yussouff free energy functional \cite{Ramakrishnan_Yussouff_1979}. Following our earlier studies, the caging potential is calculated by assuming that the cage is static when the particle moves by a small distance $\Delta r$. The mean-field caging potential is expressed in terms of the static structure factor/radial distribution function of the liquid. In previous studies, some of us have shown that the depth of caging potential is coupled to the dynamics \cite{mohit_paper1,mohit_paper2,palak_JCP_2024,palak_polydisperse_softness}. Thus, in this study, instead of dealing with the whole potential, we deal with the absolute magnitude of the depth of the caging potential as we view the depth of the caging potential as an energy barrier. Since the presence of the heavy mass particles does not change the structure of the liquid (See Fig. \ref{radial_dist_fnc_binary}, and Fig. \ref{radial_dist_fnc_quaternary} in the Appendix \ref{gr appendix}), we can treat the mass system as a binary system, and the average depth of the mean-field caging potential is written as\cite{manoj_PRL_2021,mohit_paper1,mohit_paper2,palak_JCP_2024}.
\begin{equation}
 \small \beta\Phi_{r}^{B}(\Delta r = 0) =
 -4\pi\rho \int r^2dr \sum_{i=1}^2\sum_{j=1 }^2 \chi_{i}\chi_{j} C_{ij}(r) g_{ij}(r)
 \label{binary_phir}
\end{equation}
\noindent
where the separation between the tagged particle and its neighbors is denoted by $r$. Here $\beta= 1/k_BT$, $k_B = 1$, and $\rho$ is the density of the system. The tagged particle's distance from its equilibrium position is denoted by $\Delta r$. $\chi_{i} = \frac{N_i}{N}$ is the fraction of particles in type $i$. N is the total number of particles in the system. Note that the mean-field caging potential is a function of the radial distribution function (rdf), which can be expressed as\cite{Hansen_and_McDonald}; $g_{ij}(r) =\frac{V}{N_{i}N_{j}}\Big<\sum_{\alpha=1}^{N_{i}} \sum_{\beta=1, \beta \neq \alpha}^{N_{j}} \delta (r- r_\alpha +r_\beta) \Big>$ where V is the system's volume, and $N_i$, $N_j$ are the number of particles of the $i$ and $j$ types, respectively. $r_\alpha$, $r_\beta$ are the $\alpha^{th}$ and $\beta^{th}$ particle's positions in the system, respectively. According to Hypernetted chain approximation, the direct correlation function, $C_{ij}(r)$, can be represented as a function of the interaction potential and the rdf and written as, $C_{ij}(r) = -\beta u_{ij}(r) + [g_{ij}(r)-1] - \ln[g_{ij}(r)]$ \cite{Hansen_and_McDonald}. 
The beauty of this expression is that although derived from a microscopic theory, it can also be written down intuitively by using simple physical arguments. The mean-field caging potential is a potential which a particle feels due to its neighbors. Thus, the potential is a product of the probability of finding the neighbors, given by the rdf, and the interaction between the neighbors and the particle, given by the direct correlation function.

In recent work, some of us have shown that the mean-field caging potential of the pinned system is different from that of the binary system \cite{palak_JCP_2024}. Here, we sketch the steps to arrive at the expression of the mean-field caging potential for a pinned system. First, when we pin particles in a system, we need to treat the pinned particles as a separate species. Thus, a binary system becomes a quaternary system. For a quaternary system, the caging potential is expressed as;

\begin{equation}
\beta\Phi_{r}^ {Q}(\Delta r = 0) = -4\pi\rho \int r^2dr \sum_{i=1}^4 \sum_{j=1 }^4 \chi_{i}^{'} \chi_{j}^{'} C_{ij}(r) g_{ij}(r)
\label{quaternary_phir}
\end{equation}
\noindent
where, $\chi_{i}^{'} = \frac{N_i^{'}}{N}$ is the fraction of particles in type $i$. N is the total number of particles in the system. If we assume that a fraction, $c$ of particles are pinned, then $N_{1}^{'} = (1-c)N_1, N_2^{'} = (1-c)N_2, N_3^{'} = cN_1, N_4^{'} = cN_2, \chi_{i}^{'} = \frac{N_{i}^{'}}{N}$. Here, 1 and 2 represent the type A and type B particles, which are mobile, and 3 and 4 represent type A and type B particles, which are pinned\cite{palak_JCP_2024}.

In the calculation of the average caging potential in the pinned system, we consider that only the mobile particles contribute to the total caging potential. Thus, the average caging potential of the mobile particles in the presence of the pinned particles is given by a modified quaternary system\cite{palak_JCP_2024}. Note that, as we have shown earlier, only this modified quaternary system can describe the excess entropy of the pinned system, which ultimately predicts a vanishing of the configurational entropy at a higher temperature, one of the main objectives of pinning particles \cite{palak_JCP_2024}. The expression of the caging potential in the modified quaternary system $\beta \Phi_{r}^{M}$, can be expressed as\cite{palak_JCP_2024};
\begin{equation}
\begin{split}
 \beta \Phi_{r}^ {M} (\Delta r = 0) & = -4\pi\rho \int r^2dr \sum_{i=1}^2 \chi_{i} \Big[ \sum_{j=1 }^2 \chi_{j}^{'} C_{ij}(r)g_{ij}(r) \\
 & \hspace{1.5cm} + 2 \times \sum_{j=3}^4 \chi_{j}^{'} C_{ij}(r)g_{ij}(r) \Big ]
\end{split}
 \label{modified_phir}
\end{equation}
\noindent
In the above expressions of mean-field caging potential, factor '2' appears in the terms representing the effect of the pinned particles on the mobile particles. This shows that the contribution of a pinned particle in confining (deepening the caging potential) a mobile particle is doubly stronger than another mobile particle. A detailed discussion of the modified quaternary system can be found in Reference \cite{palak_JCP_2024}. 

The above expressions represent the average caging potential of the binary, quaternary, and modified quaternary systems. However, in the calculation of the structure dynamics correlation, we are interested in the value of the same at the per particle level. 
To present cleaner statistics, we only consider the mobile ``A" particles.
The caging potential of a mobile ``A" type particle in a binary system can be expressed at the microscopic level by removing the first summation in Eq. \ref{binary_phir},

\begin{equation}
 \beta\Phi_{r}^{B}(A,\Delta r = 0) = -4\pi\rho \int r^2dr \sum_{j=1 }^2 \chi_{j} C_{1j}(r) g_{1j}(r) 
 \label{binary_phir_A}
\end{equation}

Similarly, by removing the first summation in Eq. \ref{modified_phir}, the mean-field caging potential for a mobile ``A" type particle in a modified quaternary system can be expressed as\cite{palak_JCP_2024};

\begin{equation}
\begin{split}
 \beta \Phi_{r}^ {M} (A,\Delta r = 0) & = -4\pi\rho \int r^2dr \Big [ \sum_{j=1 }^2 \chi_{j}^{'} C_{1j}(r) g_{1j}(r) \\
 & \hspace{0.8cm} + 2 \times \sum_{j=3}^4 \chi_{j}^{'} C_{1j}(r)g_{1j}(r) \Big ]
\end{split}
 \label{modified_quaternary_phir_A}
\end{equation}

In the above expression, the rdf is now calculated at per particle level, and the details of the calculation for the local mean-field caging potential are given in the Appendix \ref{mean-field caging potential appendix}.

It was shown earlier that for the pinned system, the structure dynamics correlation obtained by treating the system as a modified quaternary system was higher than that obtained by treating it as a binary system\cite{palak_JCP_2024}. We expect that in the mass system, if the heavy mass particles act as soft pinning centres, then treating the system as a modified quaternary system will lead to higher structure dynamics correlation compared to that obtained by treating the system as binary.

We compare the structure dynamics correlation by computing the Spearman rank correlation $C_{R}(X,Y) = 1 - \frac{6 \sum d_{i}^2}{m(m^2 -1)}$ where $d_{i}$ is a difference between the ranks of X and Y of each observation and m denotes the total number of observations. Like our previous studies, the rank correlation is calculated 
between the absolute value of the inverse depth of the mean-field caging potential and the isoconfigurational mobility of the lighter ``A" particles. The depth of the caging potential is calculated by treating the systems as binary and modified quaternary systems. In the latter case, the heavier particles are treated like pinned particles. 

Our earlier study shows that this present theoretical formulation works when we have a well-defined cage \cite{mohit_paper1}. The well-defined cage is present only below the onset temperature, where the short and long time dynamics are decoupled. As shown in the Appendix \ref{inherent structure}, from the inherent structure analysis, we find that the onset temperatures of all the systems are similar. Thus, we assume that the onset is around $T=0.8$, which is the onset temperature of the KA model \cite{mohit_paper1}, and the structure dynamics correlation is calculated primarily below this temperature. 

\begin{figure}[h!]
 \centering
 \includegraphics[width=0.4\textwidth]{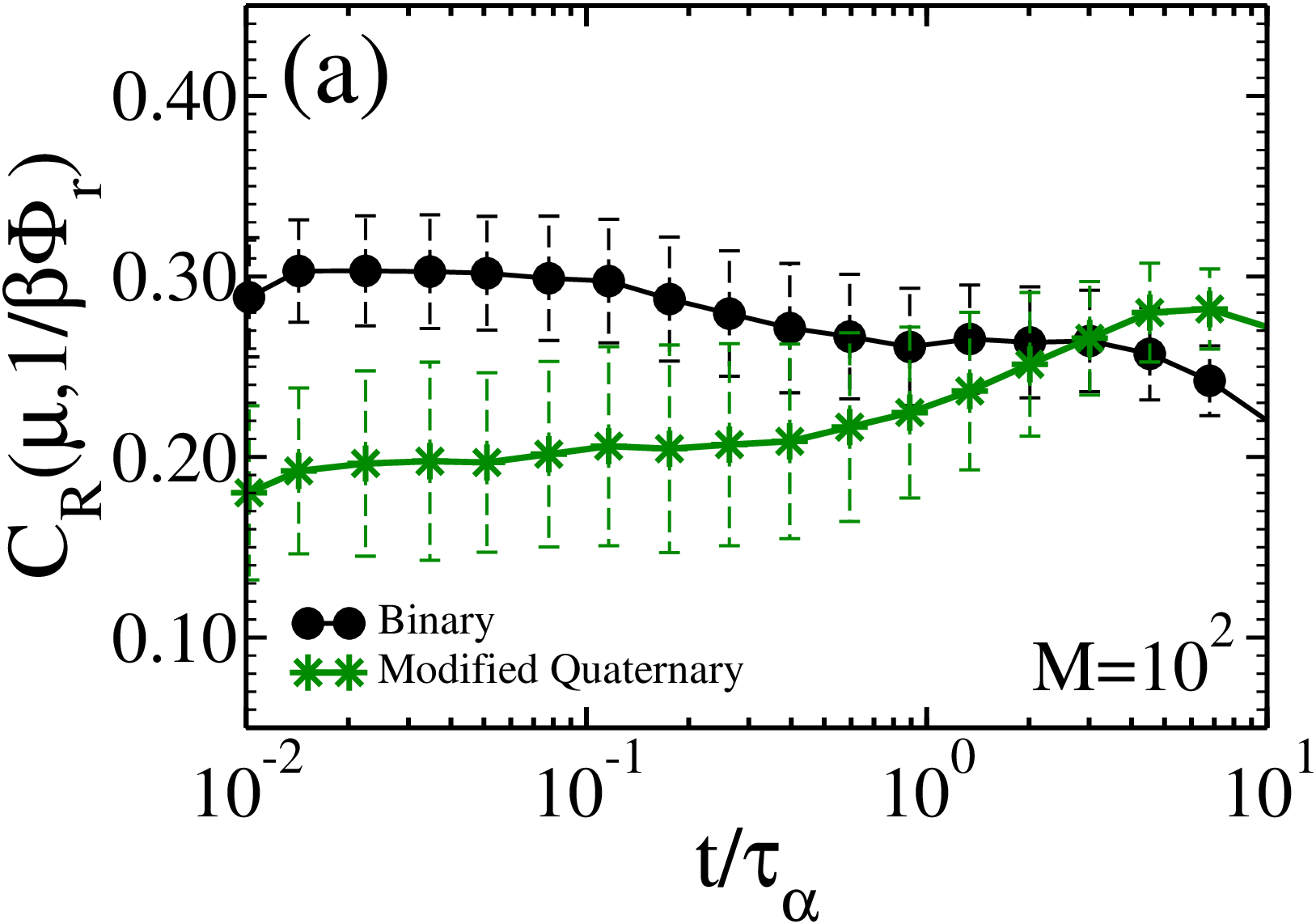}
 \\
 \vspace{0.6cm}
 \includegraphics[width=0.4\textwidth]{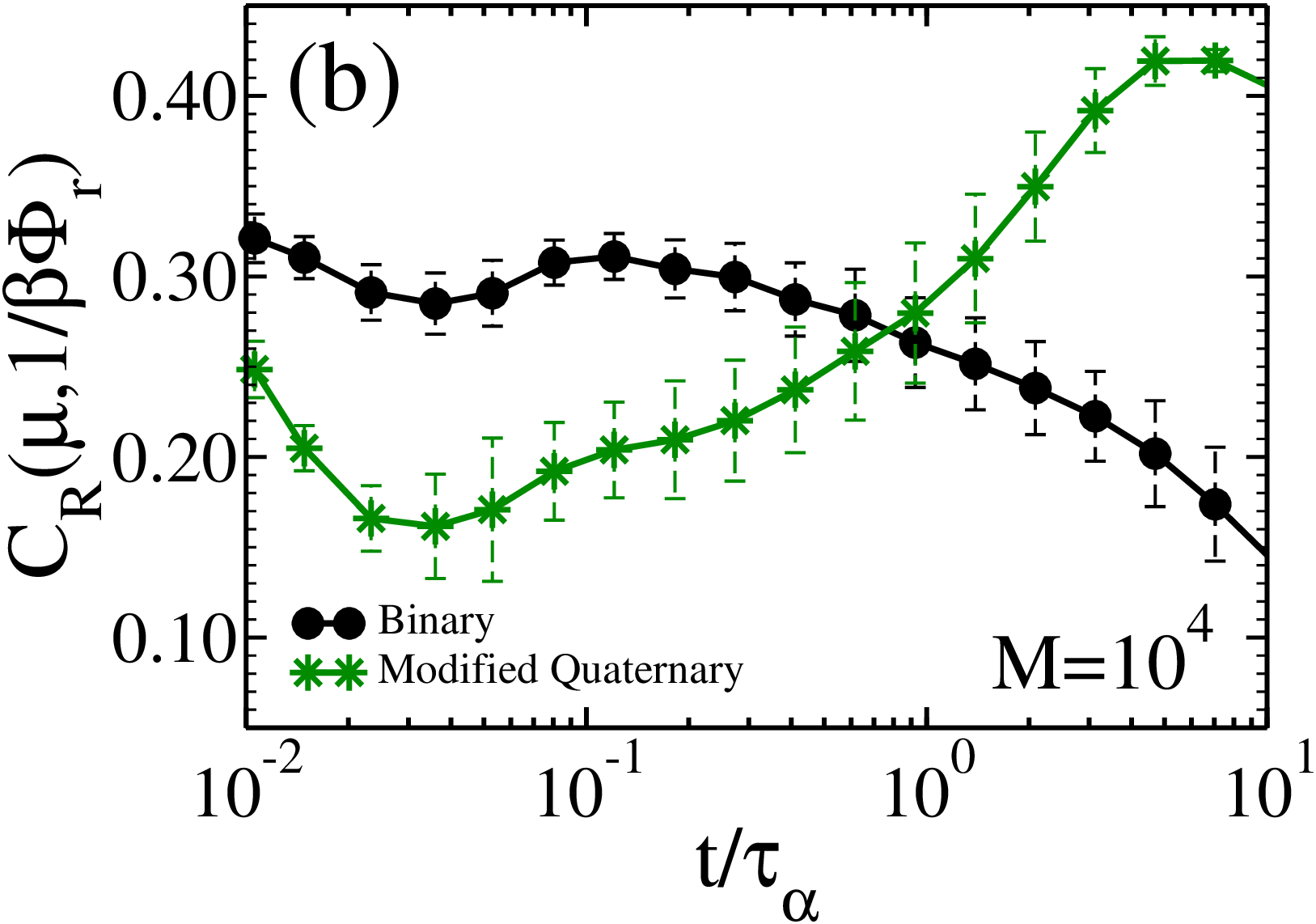}
 \\
 \vspace{0.2cm}
 \caption{The Spearman rank correlation, $C_R$, between the mobility, $\mu$, and the absolute value of the inverse depth of the caging potential,$1/\beta \Phi_{r}$, considering the heavy mass system as binary(black, circle) and modified quaternary (dark-green) plotted as a function of scaled time, $t/\tau_{\alpha}$ (where $\tau_{\alpha}$ is the relaxation time of lighter particle). (\textbf{a}) System with $M=10^{2}$ at T=0.48 and $\Delta Q$ = 0.02 (\textbf{b}) System with $M=10^{4}$ at T=0.51 and $\Delta Q$ = 0.04. Note that for each system, the analysis is done at the state point with the smallest $\Delta Q$ value studied here. Error bars are the standard deviation of all 5 sets of individual isconfiguration runs.} 
 \label{M2_M4}
 \end{figure}

In Fig. \ref{M2_M4}, we plot the Spearman rank correlation for M2 and M4 systems at low temperatures. We find that for both systems, compared to the modified quaternary framework, the system is better described by the binary framework. This is because at these temperatures, the dynamics of the heavier and lighter mass particles are coupled, as shown by the small $\Delta Q$ values, and the heavy particles cannot act as soft pinning centres. 

\begin{figure*}
\centering
\includegraphics[width=0.3\textwidth]{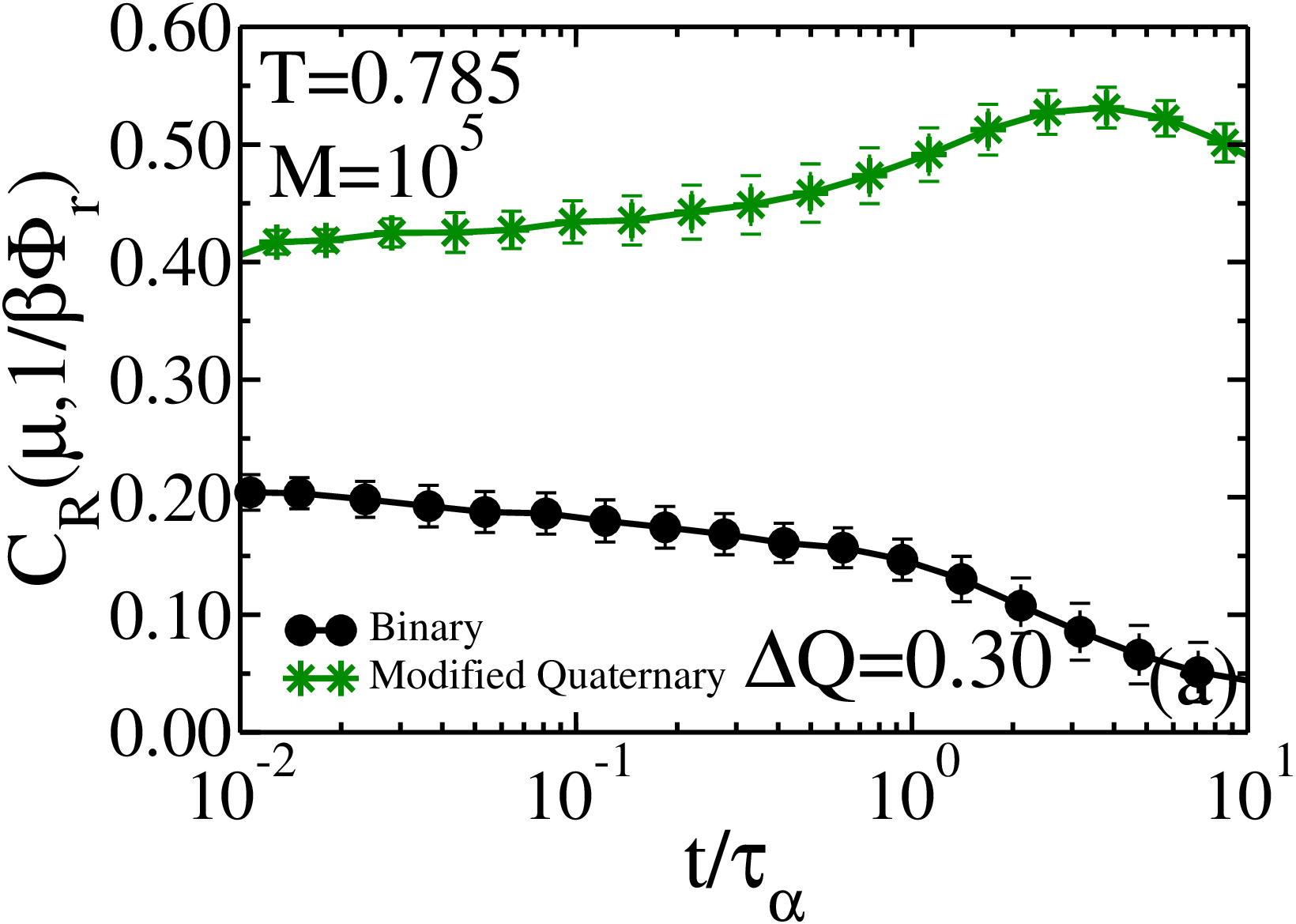}
\hspace{0.1cm}
\includegraphics[width=0.3\textwidth]{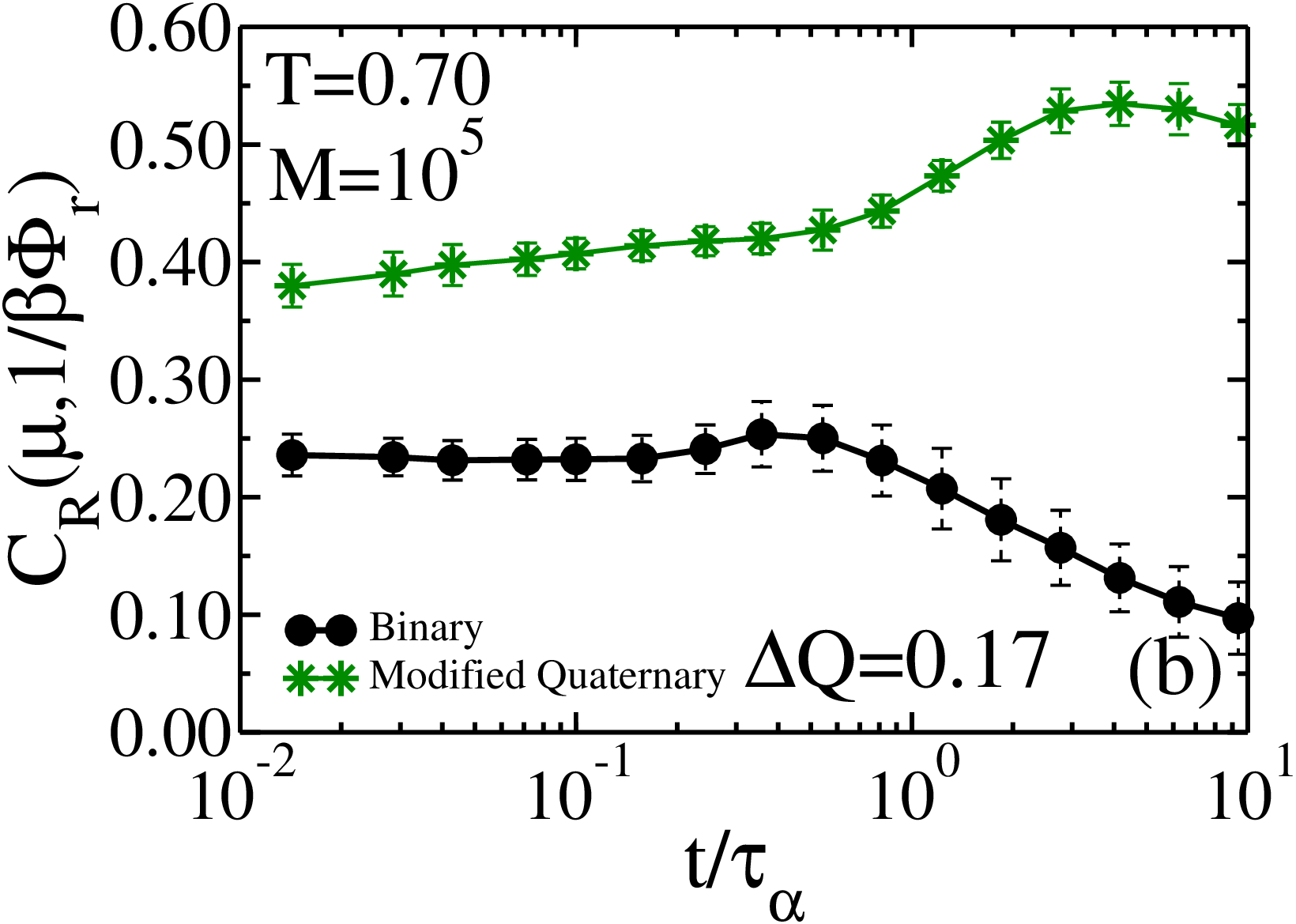}
\hspace{0.1cm}
\includegraphics[width=0.3\textwidth]{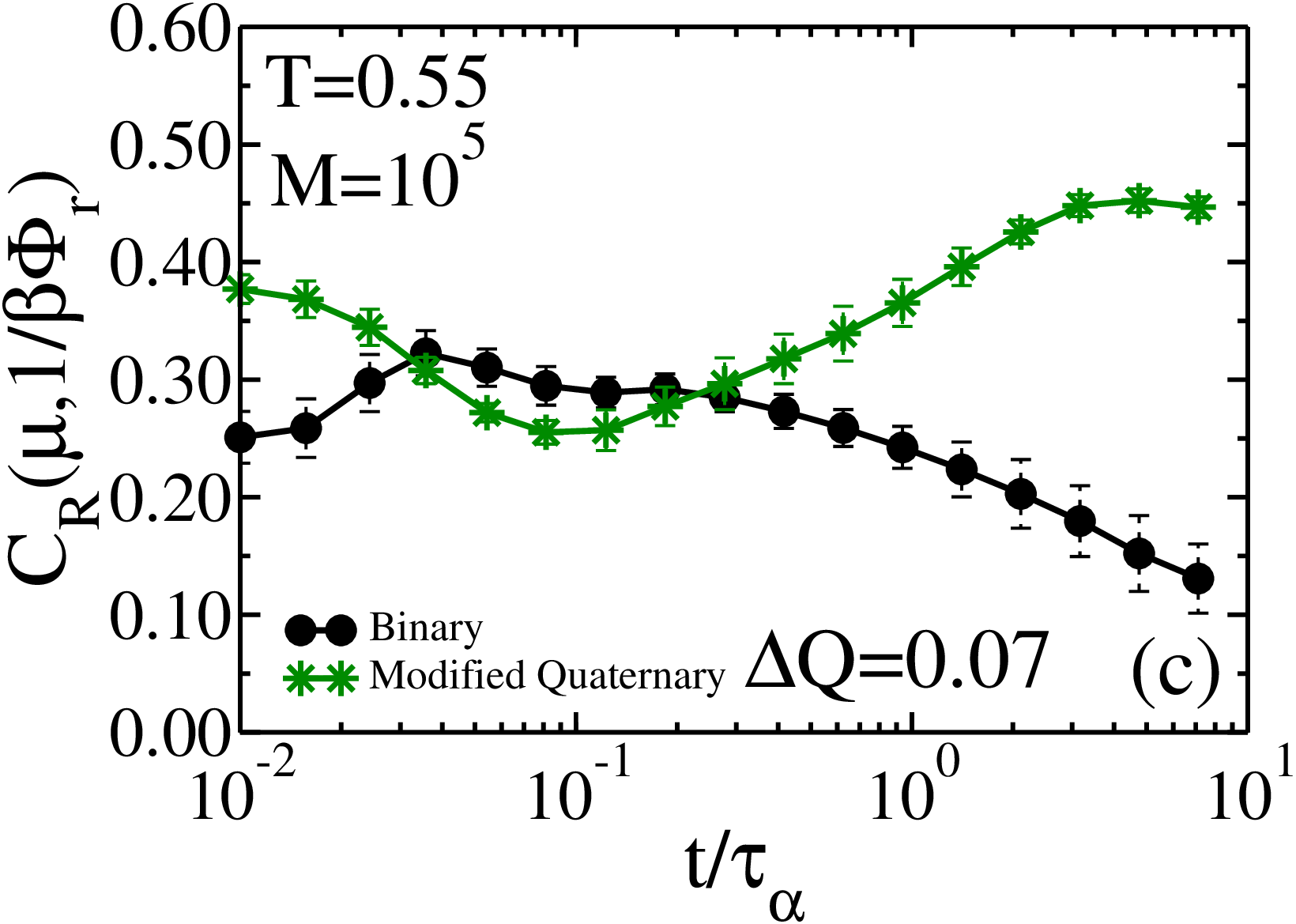}
\\
\vspace{0.4cm}
\includegraphics[width=0.3\textwidth]{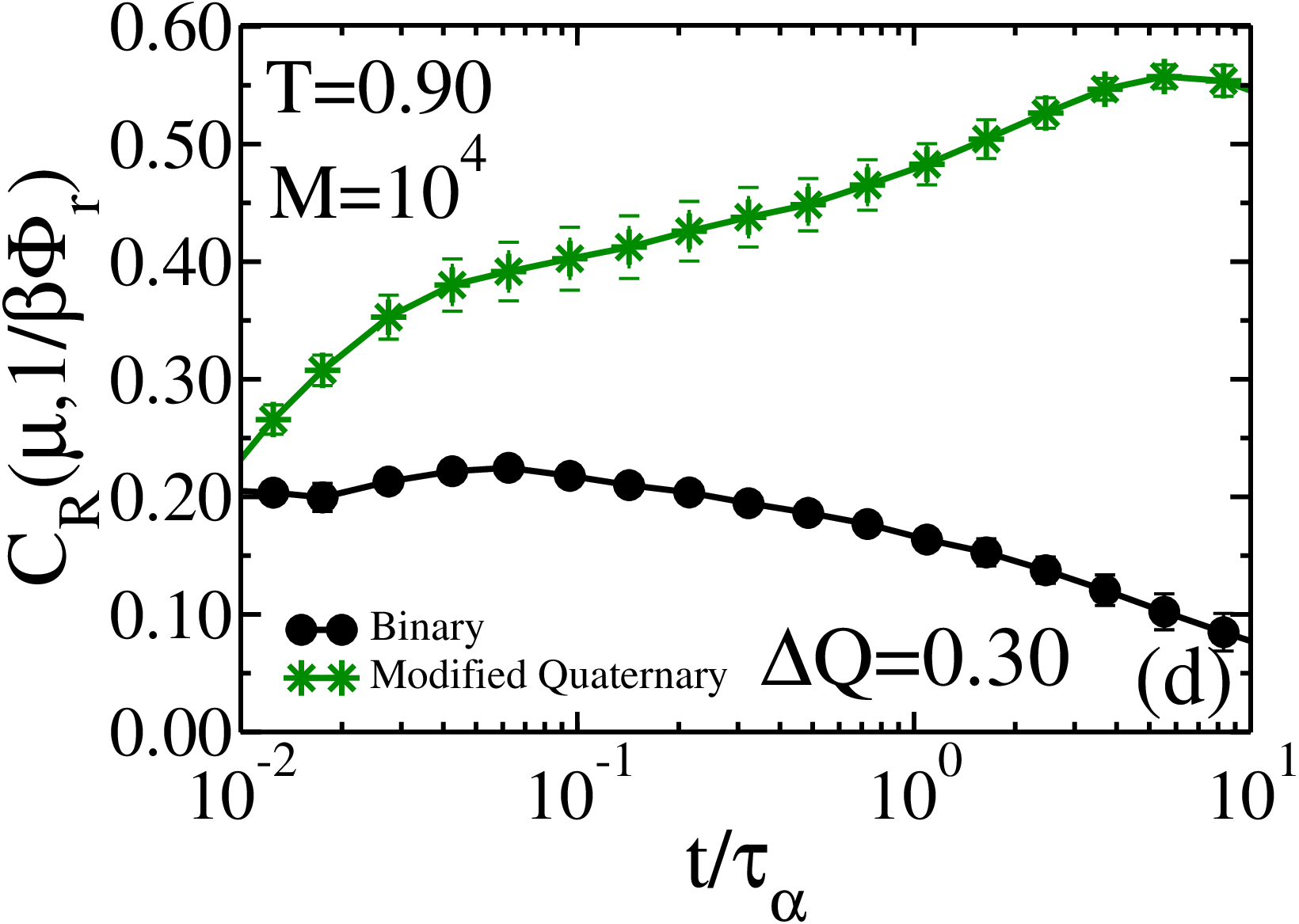}
\hspace{0.1cm}
\includegraphics[width=0.3\textwidth]{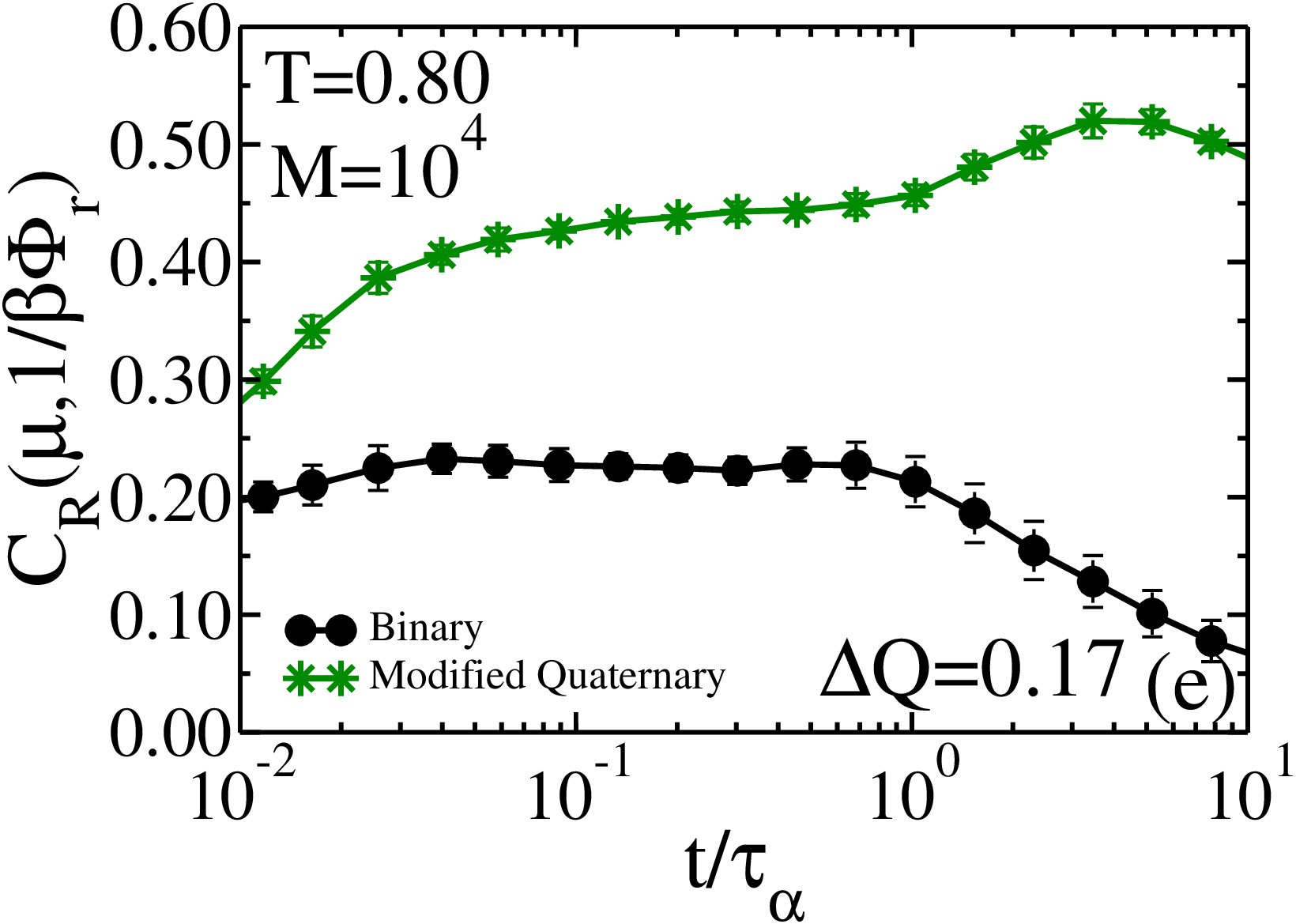}
\hspace{0.1cm}
\includegraphics[width=0.3\textwidth]{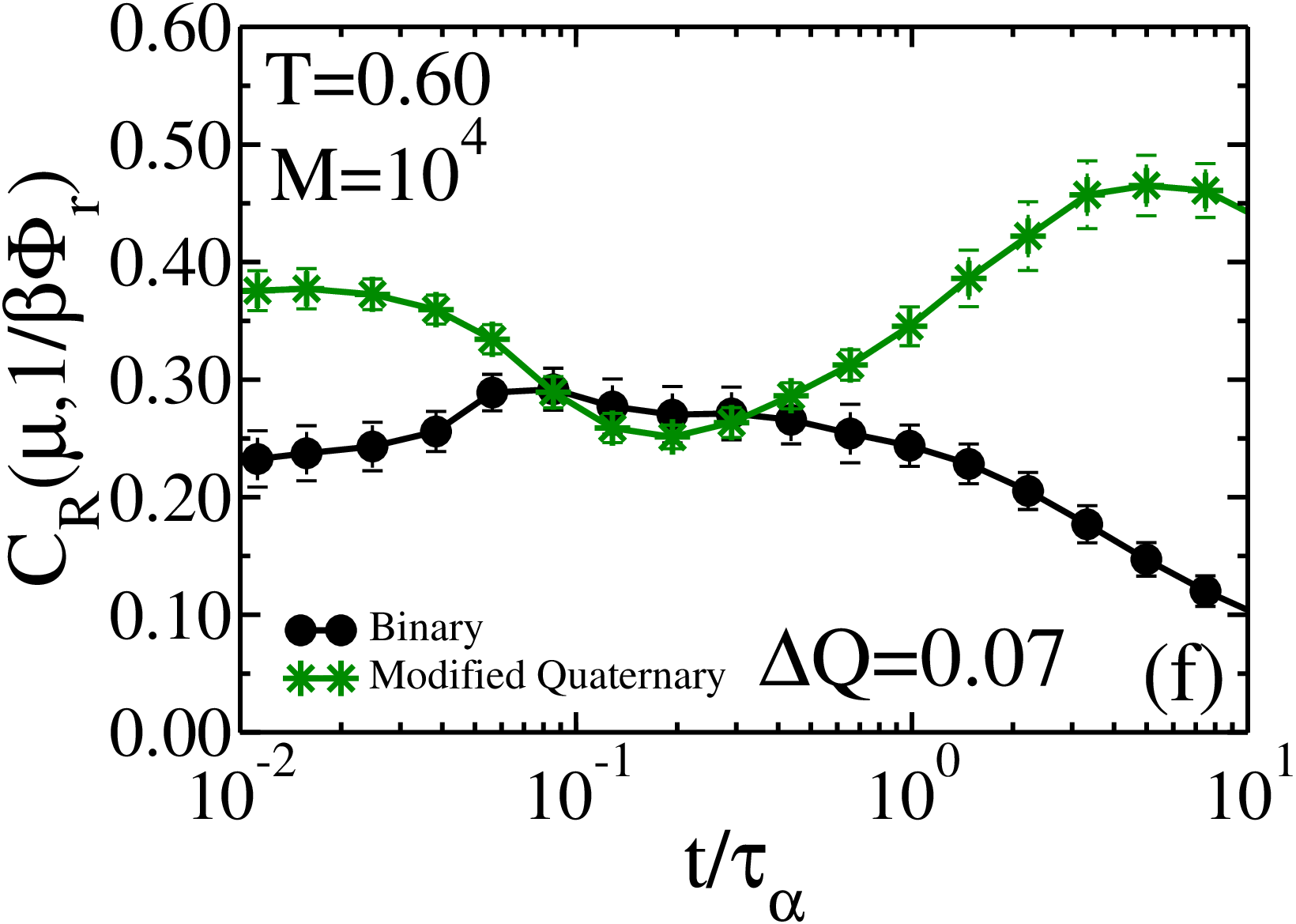}
\\
\vspace{0.2cm}
\caption{The Spearman rank correlation, $C_R$, between the mobility, $\mu$, and the absolute value of the inverse depth of the caging potential,$1/\beta \Phi_{r}$, for $M=10^4$ and $M=10^5$ systems at different temperatures, T and different $\Delta Q$ values, considering the heavy mass system as binary(black, circle) and modified quaternary (green, star) plotted as a function of scaled time, $t/\tau_{\alpha}$ (where $\tau_{\alpha}$ is the relaxation time of lighter particle). \textbf{Top panel}, $C_R$ for $M=10^5$ system (\textbf{a})T=0.785, $\Delta Q$ = 0.30 (\textbf{b})T=0.70, $\Delta Q$ = 0.17 (\textbf{c})T=0.55, $\Delta Q$ = 0.07 and \textbf{Bottom panel:}, $C_R$ for $M=10^4$ system (\textbf{d})T=0.90, $\Delta Q$ = 0.30 (\textbf{e})T=0.80, $\Delta Q$ = 0.17 (\textbf{f})T=0.60, $\Delta Q$ = 0.07. Error bars are the standard deviation of all 5 sets of individual isconfiguration runs.} 
\label{M4_M5}
\end{figure*}

For the $M5$, below onset, we can explore different ranges of $\Delta Q$ values. In Fig. \ref{M4_M5}, we show that where $\Delta Q =0.17$, the correlation for the modified quaternary system is much higher than the binary system. The difference in correlation obtained by treating the system under the modified quaternary framework and the binary framework decreases with $\Delta Q$, and eventually, at a low enough temperature, the system is better described as a binary system. In the same figure, we also plot the structure dynamics correlation of the M4 system at different temperatures but similar $\Delta Q$ values as that for the M5 system. Note that the correlation between structure and dynamics depends both on the temperature and the degree of coupling between lighter and heavier mass particles. The study shows that the degree of coupling plays a dominant role in describing the structure dynamics correlation, as for different systems at similar $\Delta Q$ values but different temperatures (albeit not drastically different), the correlation appears to be similar.

\begin{figure}
 \centering
 \includegraphics[width=0.4\textwidth]{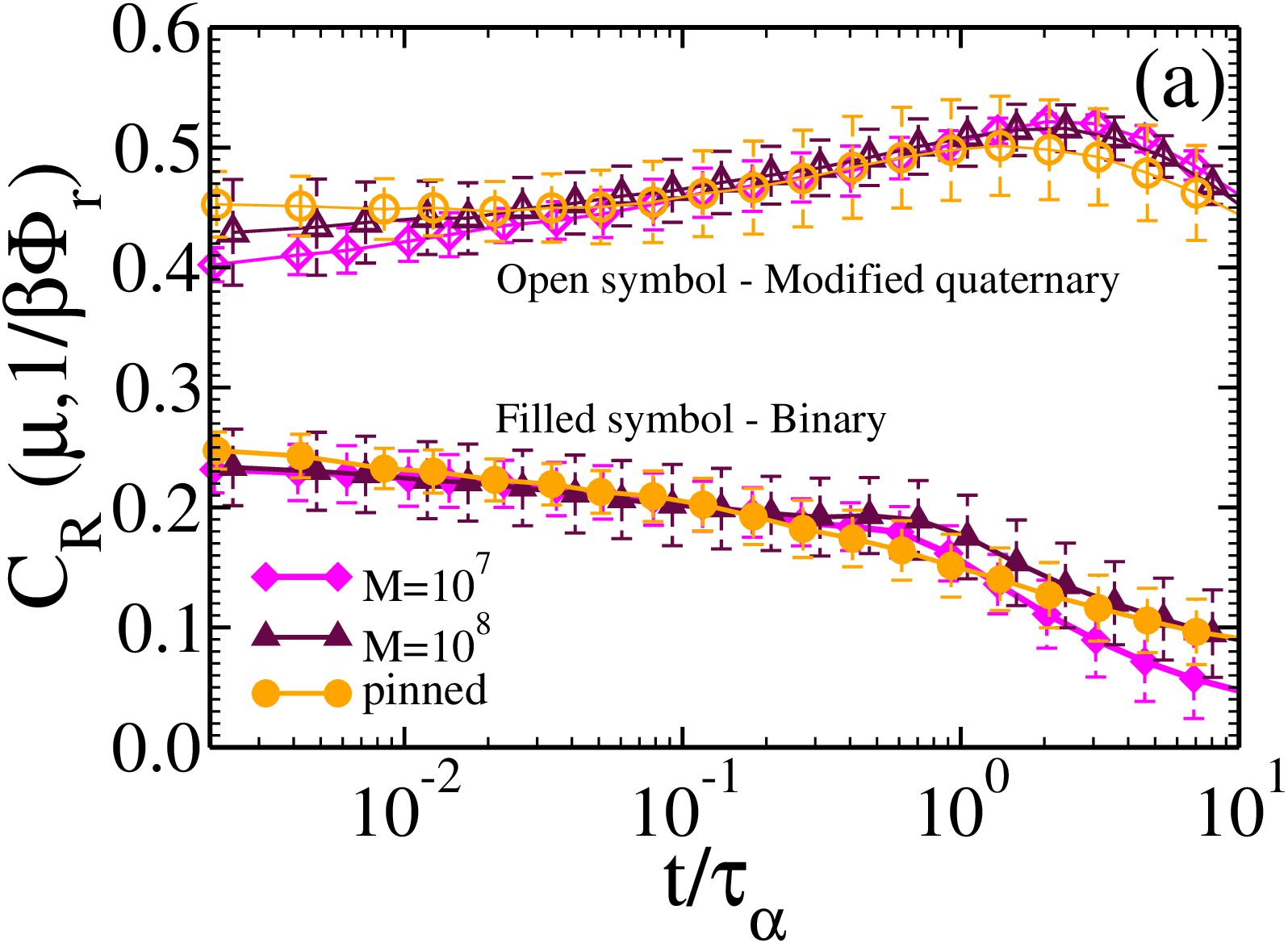}
 \\
 \vspace{0.4cm}
 \includegraphics[width=0.4\textwidth]{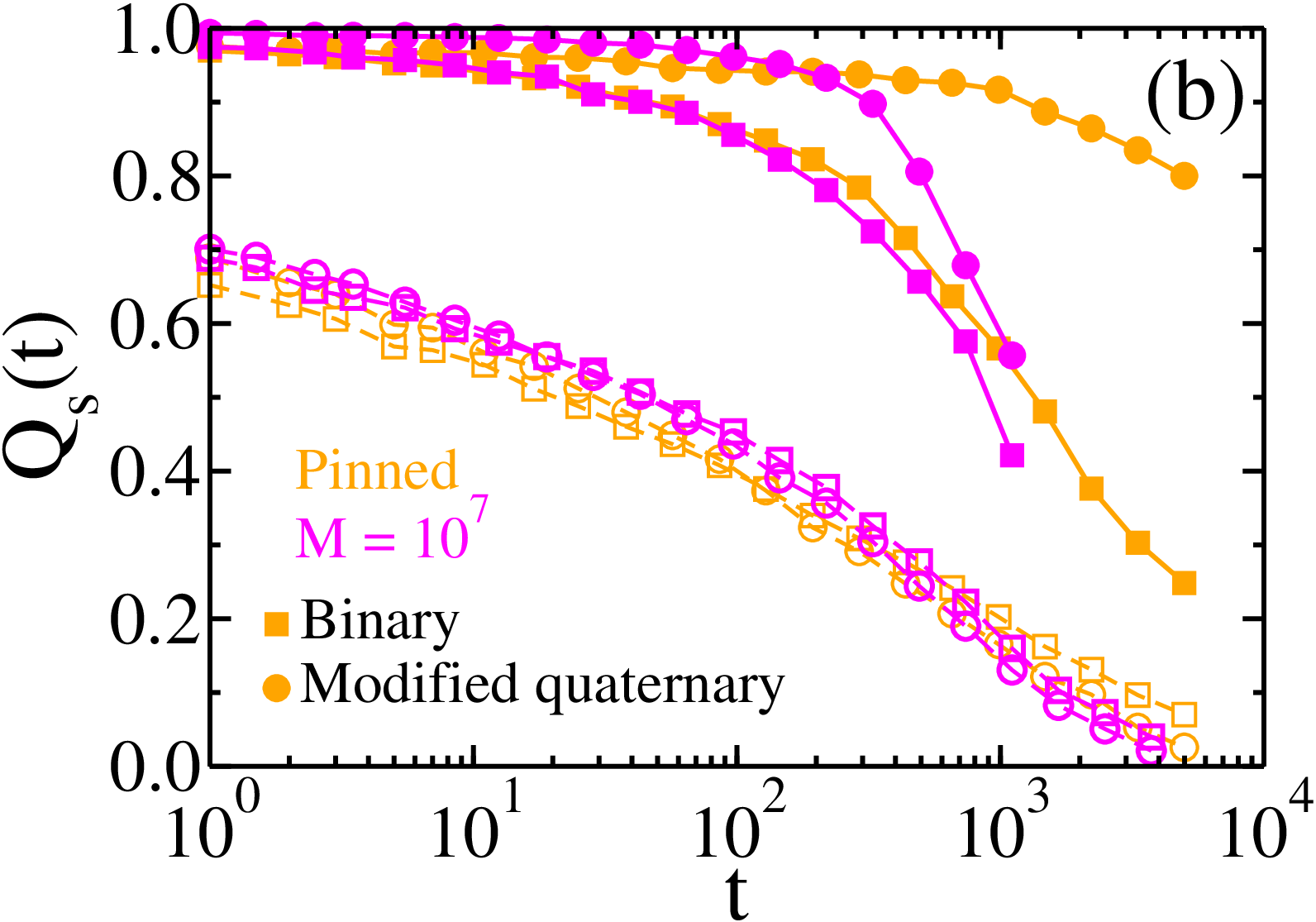}
 \\
 \vspace{0.4cm}
 \includegraphics[width=0.4\textwidth]{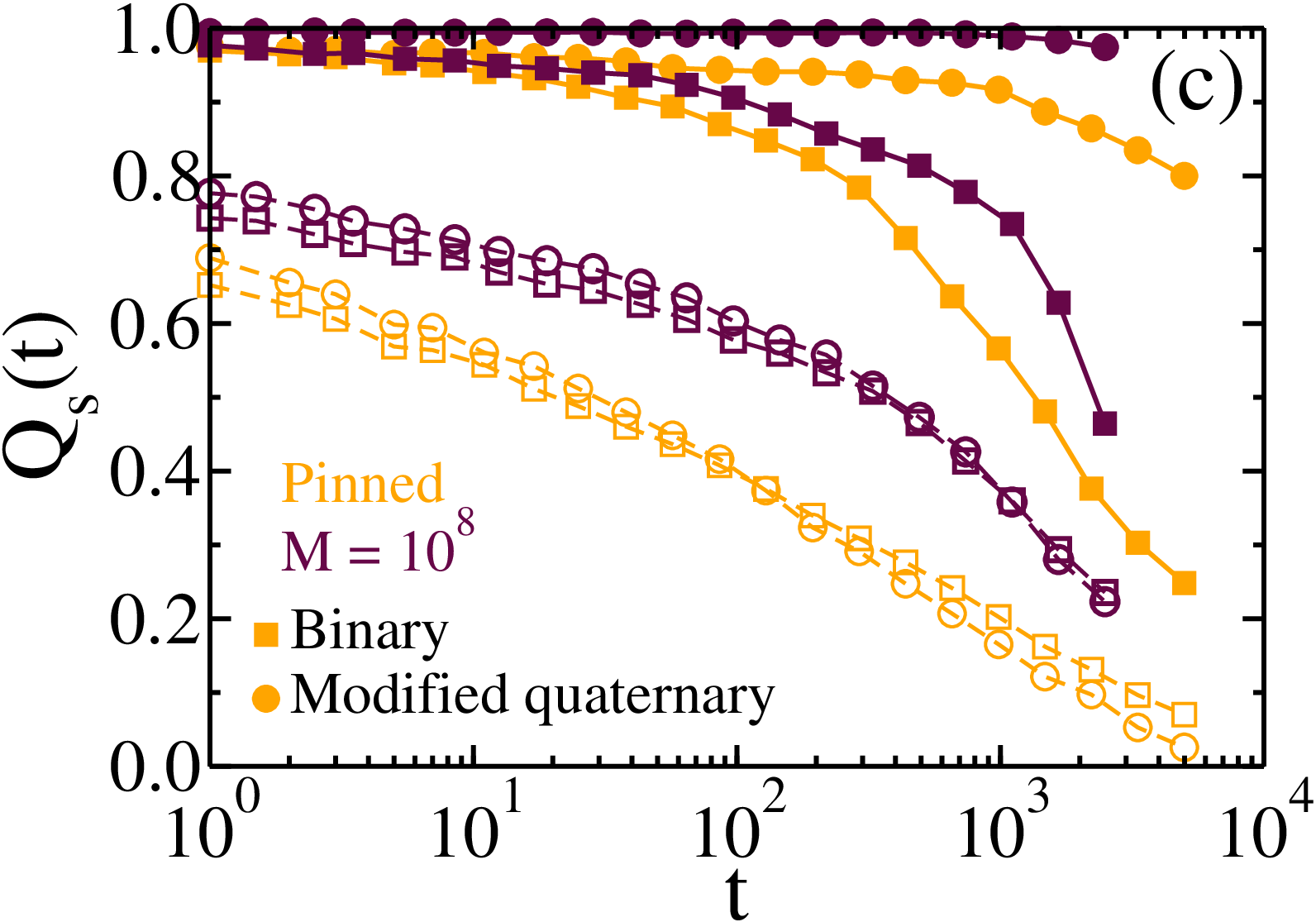}
 \\
 \vspace{0.2cm}
 \caption{\textbf{(a)}The Spearman rank correlation, $C_R$, between the mobility, $\mu$, and the absolute value of the inverse depth of caging potential,$1/\beta \Phi_{r}$, is plotted as a function of scaled time, $t/\tau_{\alpha}$ (where $\tau_{\alpha}$ is the relaxation time of the lighter mass/mobile particles in the heavy mass/pinned system) for $M=10^7$ (Magenta, diamond), $M=10^8$ (Maroon, up triangle) and pinned system(orange, circle). When the heavy mass system is treated as binary(filled symbol) and modified quaternary(open symbol). Note that the working temperature, T, is 0.65, 0.62, 0.68 for $M=10^7$, $M=10^8$, and pinned systems, respectively. Temperatures are chosen such that both the mass systems have the same value of $\Delta Q$ = 0.30, and for the pinned system, the relaxation timescale is in the same range as the mass systems. Error bars are the standard deviation of all 5 sets of individual isconfiguration runs. (\textbf{b}) - (\textbf{c}) Self overlap function, $Q_{s}^{l}$, of few ($\approx 2-3$) hard (filled symbol) and few ($\approx 2-3$) soft particles(open symbol) treating system as binary(square) and modified quaternary(circle). Note that the colour code of (\textbf{b}) and (\textbf{c}) is the same as \textbf{(a)}.}
 \label{M7_M8_PIN}
 \end{figure}

For the $M7$ system, we explore the structure dynamics correlation at $T=0.65$, which is reasonably below the onset temperature and where the $\Delta Q \approx 0.3$ (Fig. \ref{M7_M8_PIN} (a)). 
We find that there is a large difference in the structure dynamics correlation when the system is treated as a binary and modified quaternary, the latter predicting a higher correlation.
In the same figure, we also plot the structure dynamics correlation for the M8 system at $T=0.62$ and $\Delta Q \approx 0.3$. We find that, as discussed above, the behavior is similar for both systems. For comparison, we also plot the structure dynamics correlation for the pinned system at $T=0.68$, where the dynamics of the pinned system is similar to that of the lighter particles in the M7 and M8 systems. We find that the correlation obtained for the pinned system matches with the $M7$ and $M8$ systems. This shows that even at temperatures where the dynamics of the lighter and heavier particles start coupling, and $\Delta Q$ starts dropping from its high-temperature values, depending on the degree of coupling, the system may act as a soft pinning model. 

We also perform a comparative study of the dynamics of a few hardest, about 2-3 (sitting in the deepest part of the mean-field potential), and a few softest, about 2-3 (sitting in the most shallow part of the mean-field potential) particles. The farther the separation between the dynamics of the hardest and softest particles, the better the predictive power of the structural order parameter. Note that the identity of the few hardest and softest particles changes when the caging potential is calculated, treating the system as binary and modified quaternary. As shown in Fig. \ref{M7_M8_PIN} (b) and (c), for the M7, M8, and pinned system, the dynamics of the few hardest and softest particles are more apart when the systems are treated as modified quaternary systems. We find that the difference primarily comes in the dynamics of the hardest particles. In the modified quaternary system, because of the stronger confining effect of the heavy/ pinned particles, the particles which are hardest always have heavy/pinned particles as their neighbors, and due to the presence of these heavy/pinned particles, the dynamics of the hardest particles are also slower. Although the Spearman rank correlation of the M7 and M8 systems are similar from Fig. \ref{M7_M8_PIN} (b) and (c), we find that the difference in the dynamics of the hardest particles when the system is treated as binary and modified quaternary is more for the M8 system. Thus, as expected with the increase in the mass of the heavy particles, the system is better described under the modified quaternary framework. As discussed earlier, with an increase in mass of the heavier particles, the coupling between the dynamics of the heavy and light particles, even at low temperatures, becomes weak, and the $\Delta Q$ saturates at a finite value (Fig. \ref{deltaq}); thus, the system becomes better suited to act as a soft pinning model over a wider temperature range.

\section{Ternary system and its soft pinning effect}
\label{Ternary system and its soft pinning effect}
\begin{figure}[h!]
 \centering
 \includegraphics[width=0.4\textwidth]{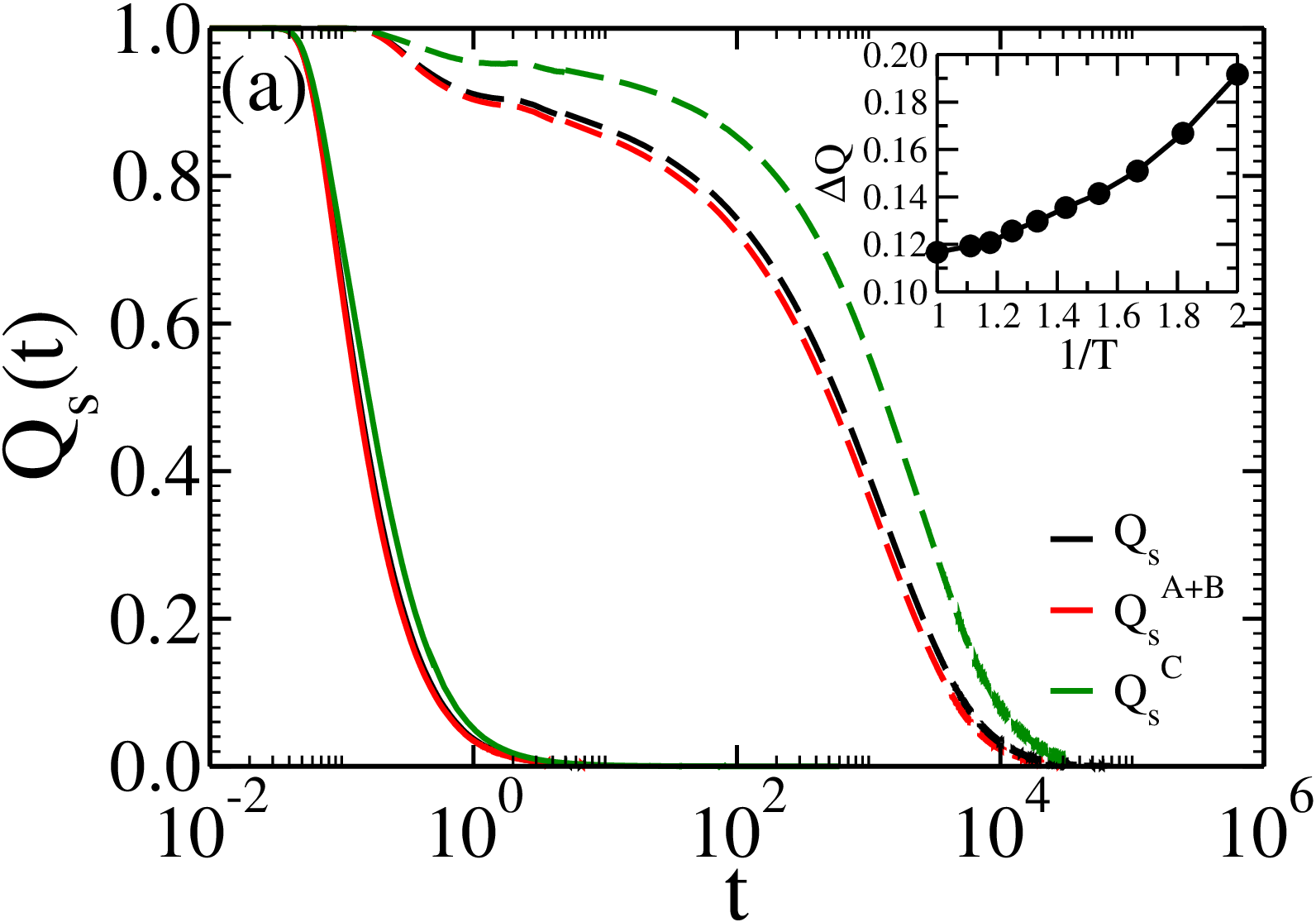}
 \\
 \vspace{0.4cm}
 \includegraphics[width=0.4\textwidth]{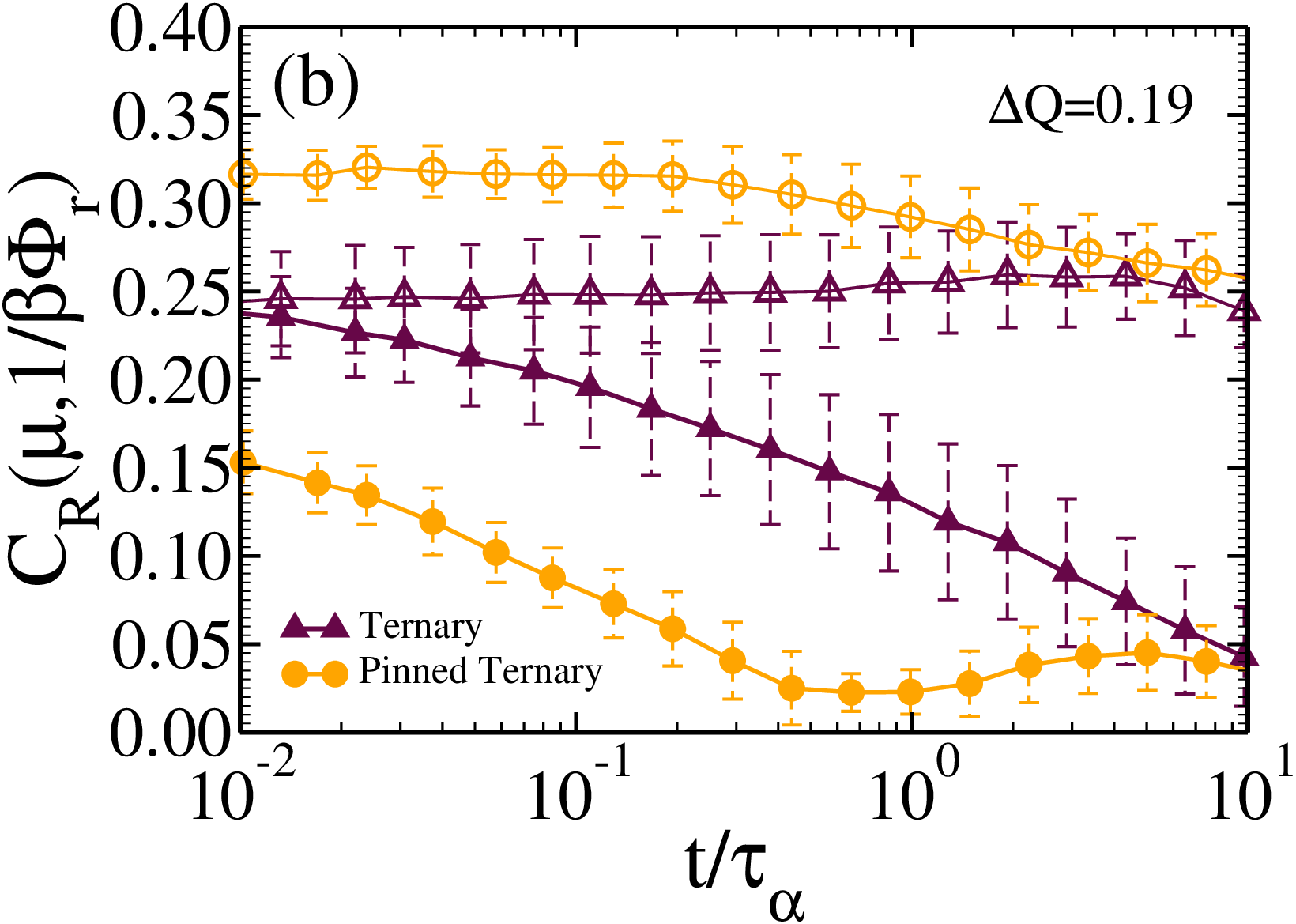}
 \\
 \vspace{0.2cm}
 \caption{(a)A comparison of the self overlap functions of the biggest``C" particles, $Q_s^{C}$, of the "A" and``B" particles,$Q_s^{A+B}$, and of the total system, $Q_{s}$. The solid lines are at a higher temperature, T=5.0 and dashed lines are at a lower temperature T=0.50; Inset- The temperature dependence of $\Delta Q =Q_s^{C}(t=\tau_{\alpha})-Q_s^{A+B}(t=\tau_{\alpha})$, where $\tau_\alpha$ is the relaxation time where the total self overlap $Q_{s} (t=\tau_{\alpha})=1/e$ (b) Spearman rank correlation, $C_R$ between the mobility $\mu$ and the absolute value of the inverse of depth of caging potential, $1/\beta\Phi_r$ for the``A" and ``B" particles in the ternary system (Maroon, triangle) at T=0.50 and in the pinned ternary system where the ``C" type of particles are pinned (orange, circle) at T=0.67. The Spearmann rank correlation is calculated considering the system as ternary, Eq. \ref{ternary_phir_A} (Filled symbol) and modified ternary, Eq. \ref{modified_ternary_phir_A} (Open, symbol). The temperatures for the systems are such chosen such that the relaxation times are similar. } 
 
 \label{ternary}
 \end{figure}

Note that the dynamics of a particle is a function of mass and size. Thus, it is possible to have soft pinning centres by also introducing a bigger size particle \cite{smarajit_soft_pinning_pnas_2024}. We study the correlation of structure and dynamics in the ternary system, where a larger C type of particle is introduced in the binary KA model. In Fig. \ref{ternary}(a), we plot the self overlap of the biggest particle and the rest of the system. We also plot the $\Delta Q =Q_s^{C}(t=\tau_{\alpha})-Q_s^{A+B}(t=\tau_{\alpha})$ in the inset, where $Q_s^{C}$ is the self-overlap function of the biggest particle and $Q_s^{A+B}$ is that of the rest of the system. $\tau_\alpha$ is the relaxation time where the total self overlap $Q_{s}(t=\tau_{\alpha})=1/e$. The difference between the mass and ternary systems is that, unlike the mass system, the decoupling between the dynamics of the A and B particles from that of the C particle, given by $\Delta Q$, increases with a decrease in temperature. We study the correlation between structure and dynamics by treating the system as ternary and also by treating the system in a way that the `C' type of particles appear pinned (modified ternary). The expressions for the depth of the caging potentials are presented in Appendix \ref{mean-field caging potential appendix}. The results are presented at the lowest temperature studied here where $\Delta Q=0.19$. We find that assuming the C type of particle as pinned gives a higher structure dynamics correlation.
However, a small $\Delta Q$ value leads to a weaker enhancement of correlation for the modified ternary system.
We also compare the results with that of a system where the `C' type of particles are pinned in the simulations. We find that, in that case, the enhancement of correlation by treating the system as a modified ternary rather than a regular ternary system is pronounced. 

\section{Conclusion}
\label{conclusion}
 The present work aims to study the behavior of multicomponent systems, particularly the impact of a small percentage ($15\%$) of heavy mass particles on the dynamics of lighter particles and how they compare to the pinned system. It has been observed that the heavy mass particles have two distinct effects on the dynamics of the lighter particles. At higher temperatures, the lack of mobility of the heavy particles hinders the motion of the lighter particles by excluding certain volumes around them, resulting in a slowdown of the dynamics, particularly affecting the collective part more than the self part. However, at lower temperatures, the dynamics of the lighter and heavier particles are coupled, and the mass of the heavy particles also influences the dynamics of the lighter particles. It is important to note that this second effect is not present in the pinned system. Therefore, at high and low temperatures, the heavier particles have different effects on the dynamics of the lighter particles, impacting both the self and collective parts differently and leading to a complex landscape of lighter particle dynamics.

Our study reveals that even in the region where the dynamics of the heavy and light mass particles are coupled, with the increase in mass of the heavy particles, this coupling becomes weaker. Thus, the effect of soft pinning of the heavy particles depends on the temperature range and on the mass of the heavy particles. 
To further understand the soft pinning effect of the heavier particles, we study the structure dynamics correlation of the lighter particles in the presence of the heavier particles where the structure is described by the mean-field caging potential \cite{manoj_PRL_2021,mohit_paper1,palak_JCP_2024}. Since the presence of heavy mass particles does not affect the structure of the system, the system can be treated as a binary system. However, in a recent study, we have shown that when we pin certain particles in the system, the mean-field caging potential changes \cite{palak_JCP_2024}. The binary system with pinned particles needs to be treated as a modified quaternary system where, compared to other mobile particles, the pinned particles have a stronger confining effect on the mobile particles by increasing the depth of the mean-field caging potential of the mobile particles. With this modification in the mean-field caging potential, we found an increase in the structure dynamics correlation in the pinned system \cite{palak_JCP_2024}. Here, we argue that when the heavy mass particles act as soft pinning centres, we should see a similar effect. The structure dynamics correlation obtained under the modified quaternary framework should be higher than that obtained by the binary framework. We study the structure dynamics correlation of the lighter particles by changing the mass of heavy particles. We show that comparing the structure dynamics correlation when the system is treated under the modified quaternary framework and binary framework allows us to identify the pinning effect of the heavy mass particles.
We also show that a similar soft pinning effect can be obtained by introducing a few big particles in the system.

This study not only explores the properties of soft pinned systems but also helps us to understand certain observations in the pinned system. For the pinned system, since the collective dynamics does not decay completely thus, its connection with the thermodynamics was debated \cite{reply_by_chandan_dasgupata,reply_by_kob}. In the present study for the mass system, although we do not connect the dynamics to the thermodynamics, we show that for high values of the heavy mass particles, the self and the collective dynamics of the lighter particles do not follow each other. This effect is similar to that observed in the pinned system \cite{reply_by_kob} and different from that observed in KA model. This hints that if we have immobile/ slow moving particles in the system, the self and collective dynamics will behave differently, and the latter will be controlled by thermodynamics. We also believe that the understanding of the present study can be applied to a large class of biological systems and solutions which are inherently multicomponent in nature and where the components relax at different timescales, giving rise to a soft pinning effect\cite{copling_plasma_2017,Cytoskeletal_Pinning_2015,soft_pinning_of_liquid_2015,interleaflet_coupling_2011}

\appendix
\numberwithin{equation}{section}

\section{ ISOCONFIGURATION RUN (IC)}
\label{Isoconfiguration run}
Harrowell and coworkers \cite{Cooper_IC_2004, Cooper_IC_2006, Harowell, Bertheir_IC} developed the powerful IC technique to look into how structure affects the dynamical heterogeneity of the particles. Among other factors, a particle's displacement can be affected by its initial momenta and structure. This technique was proposed to remove the uninteresting variation in the particle displacements arising from the choice of initial momenta and to obtain the role of the initial configuration on the dynamics and its heterogeneity. We perform five separate isoconfigurational runs for 4000 particles for each system. To ensure that all configurations are different, we start with five different high-temperature, KA configurations and cool them individually. All these high-temperature configurations are chosen such that the two sets are greater than 100$\tau_{\alpha}$ apart. To generate the heavy mass system, we use five sets of KA system configurations and randomly assign $15\%$ particles with heavy mass. Note that after this, to equilibrate the position of the particles, we run the system for 100$\tau_{\alpha}$ and then consider that as our initial configuration. These five IC have different initial structures. For each configuration, we run $N_{IC}=100$ trajectories with random starting velocities chosen from the Maxwell-Boltzmann distribution for the corresponding temperatures.

Mobility, $\mu$ is the average displacement of each particle over these 100 runs and is calculated as\cite{Harowell},

\begin{equation}
\mu^{j}(t) = \frac{1}{N_{IC}}\sum_{i=1}^{N_{IC}} \sqrt{(r_{i}^{j}(t) - r_{i}^{j}(0))^{2}}
\end{equation}
\noindent
where, at time $t$, $j^{th}$ particle's mobility is represented by the term $\mu^j(t)$. At time $t$, the position of the $j^{th}$ particle in the $i^{th}$ trajectory is denoted by the term $r_{i}^{j}(t)$, and its initial position is denoted by the term $r_{i}^{j}(0)$. We determine the average displacement or mobility for the $j^{th}$ particle at time $t$ by averaging these displacements over all isoconfigurational trajectories, $N_{IC}$.\\

\section{RADIAL DISTRIBUTION FUNCTION}
\label{gr appendix}
\begin{figure}[h!]
 \centering
 \includegraphics[width=0.23\textwidth]{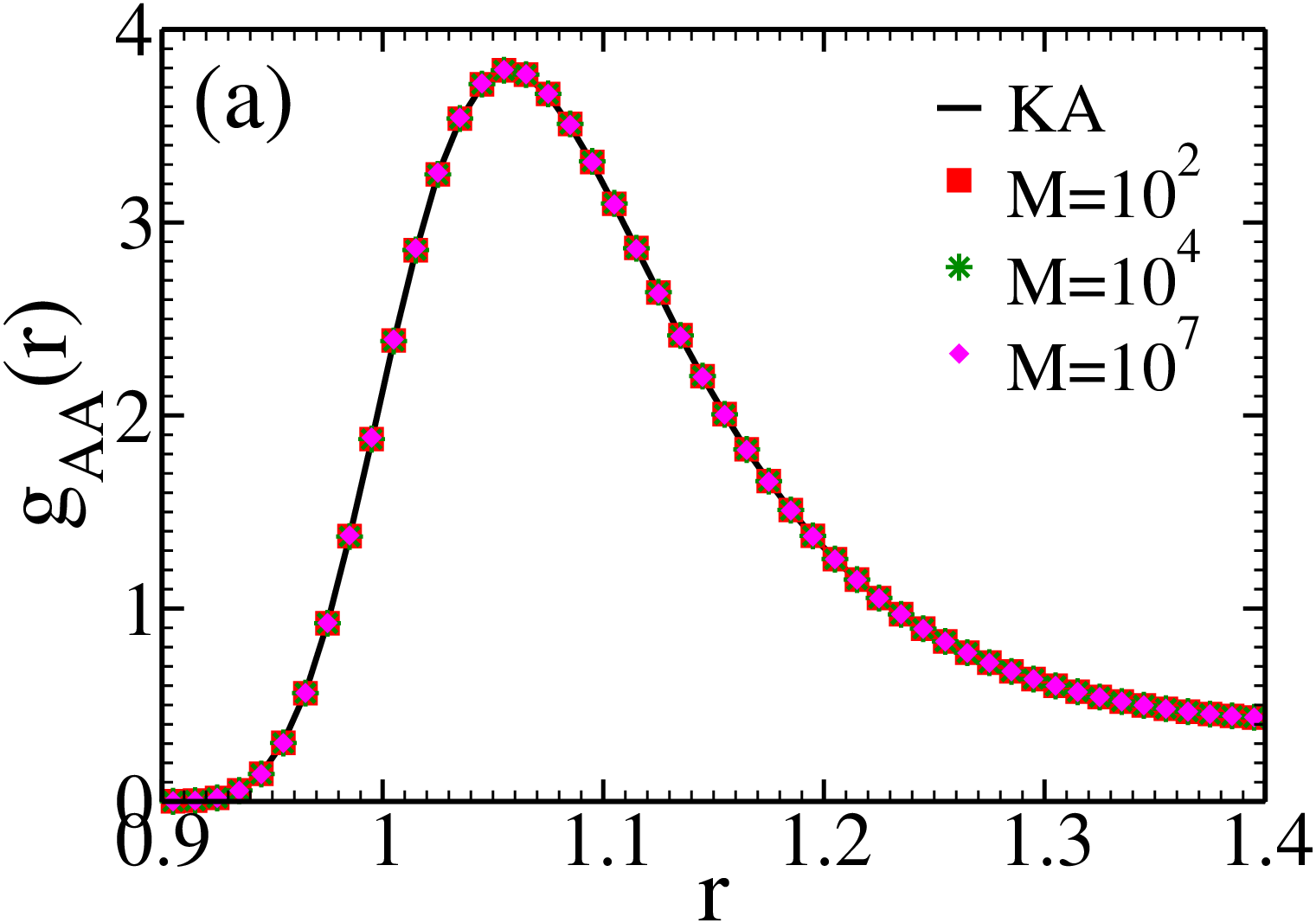}
 \hspace{0.15cm}
 \includegraphics[width=0.23\textwidth]{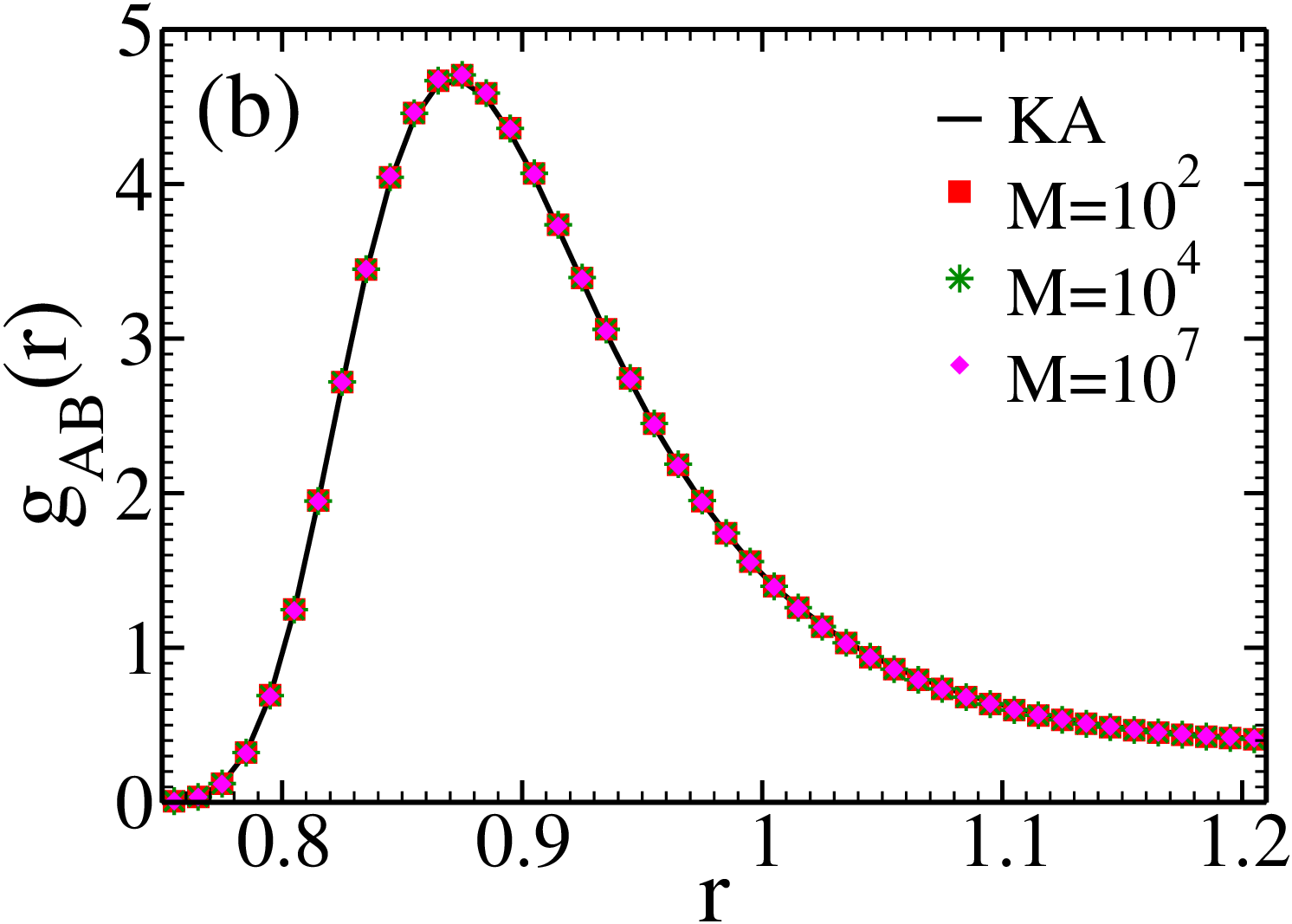}
 \\
 \vspace{0.2cm}
 \caption{Radial distribution function, g(r) while treating the heavy mass system (KA(black line), M = $10^2$ (red, square), M = $10^4$ (green, star), M = $10^7$ (magenta, diamond)) as a binary system, at T =0.65. (a)$g_{AA}$ as a function of r (b) $g_{AB}$ as a function of r. Here, A and B represent the bigger and smaller sizes of particles in the system, respectively.} 
 \label{radial_dist_fnc_binary} 
\end{figure}

Note that the mean-field caging potential is a function of the partial radial distribution function. The partial rdfs of the system where the heavy mass and lighter mass particles are not distinguished as separate species are plotted in Fig. \ref{radial_dist_fnc_binary}. 
For comparison, the partial rdf of the KA model is also plotted. As shown in the Fig. \ref{radial_dist_fnc_binary}, there is no change in the structure of the mass system.

\begin{figure}[h!]
 \centering
 \includegraphics[width=0.23\textwidth]{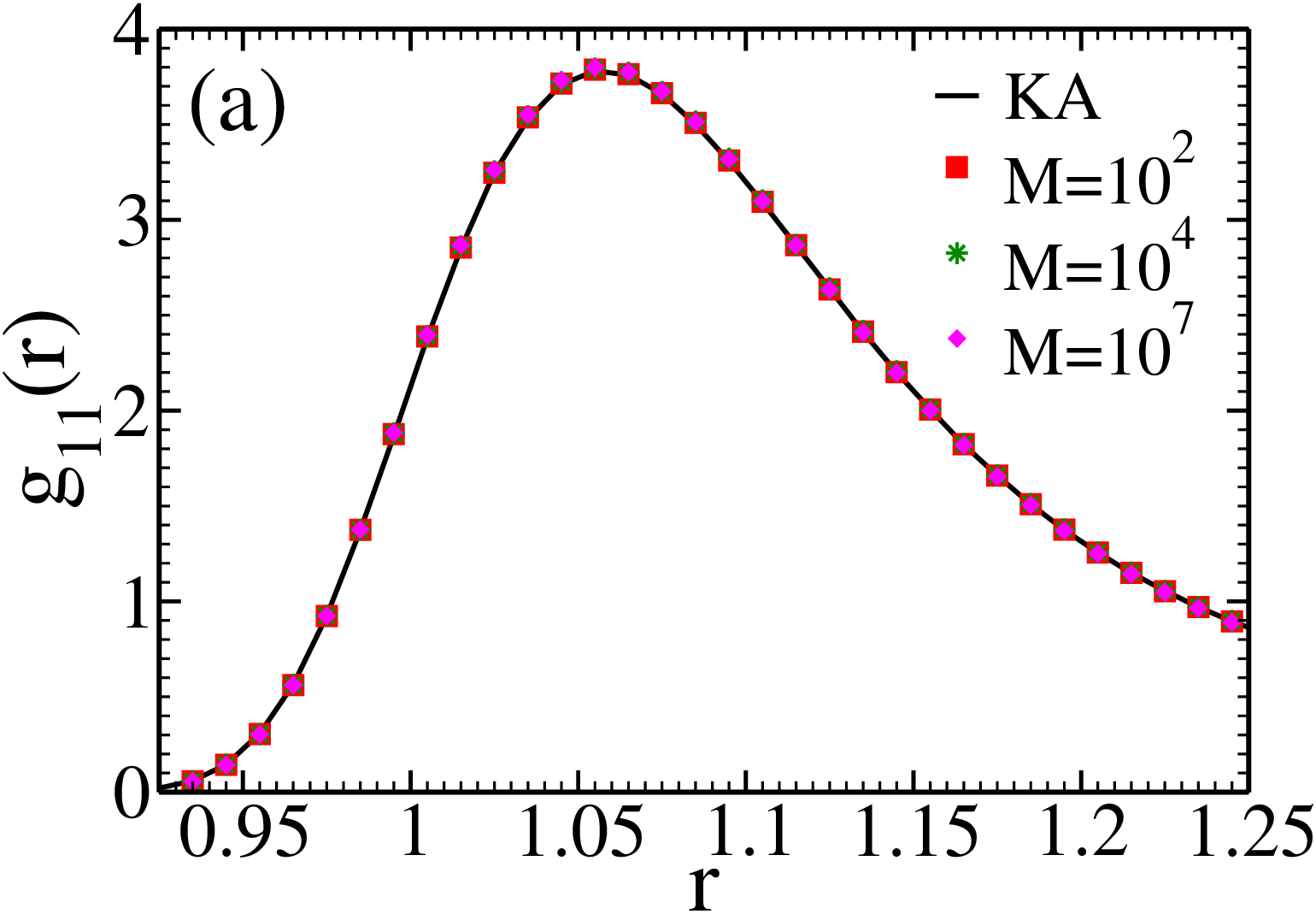}
 \hspace{0.15cm}
 \includegraphics[width=0.23\textwidth]{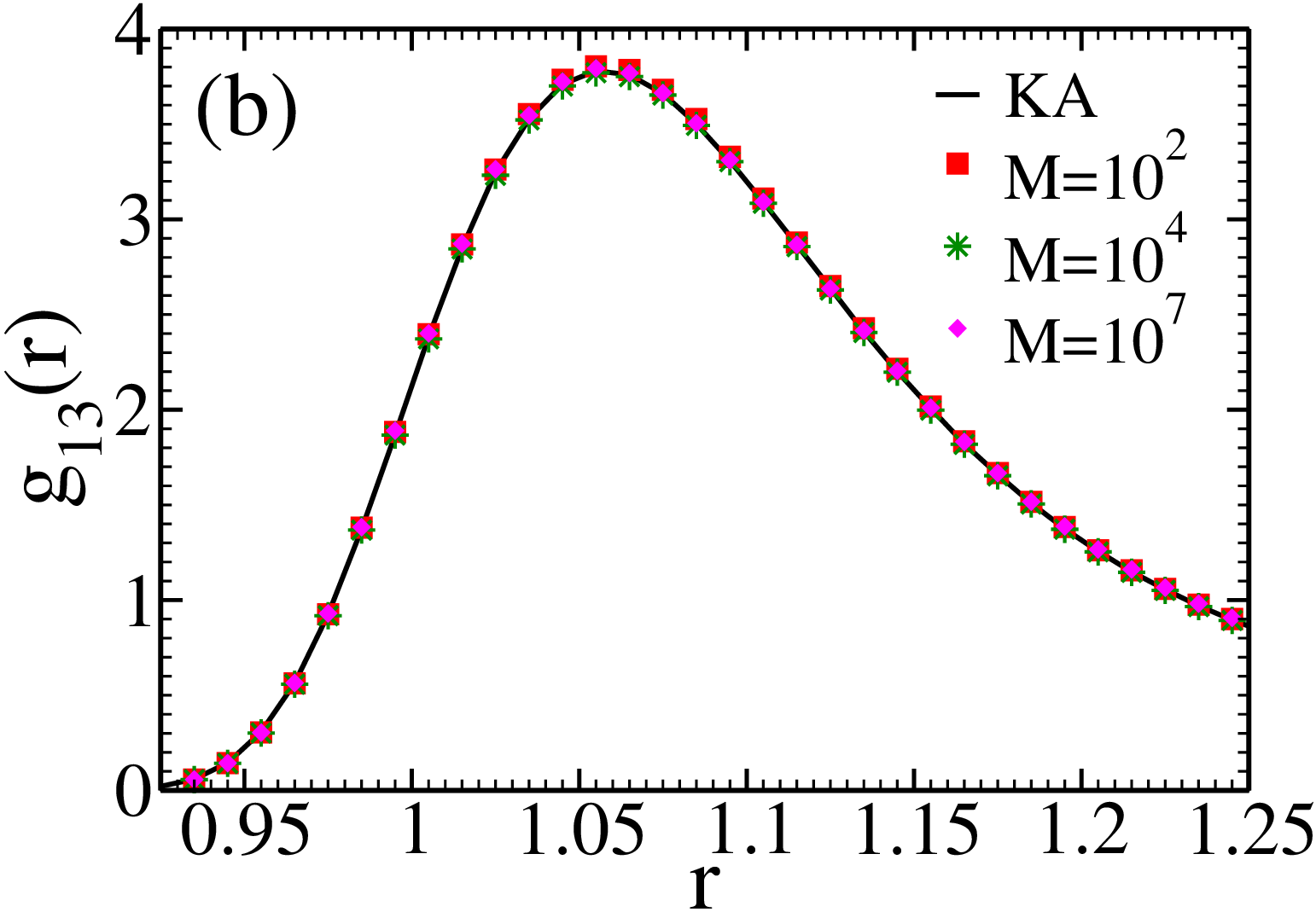}
 \\
 \vspace{0.2cm}
 \caption{Radial distribution function, g(r) while treating the heavy mass system (KA(black, line), M = $10^2$ (red square), M = $10^4$ (green, star), M = $10^7$ (magenta, diamond)) as a quaternary system, at T=0.65. (a)$g_{11}$ as a function of r (b) $g_{13}$ as a function of r. Here, we refer to the A type of particles with mass unity as 1 and A type of particles with heavy mass as 3.} 
 \label{radial_dist_fnc_quaternary} 
\end{figure}

Next, we assume that the heavy mass particles in Fig. \ref{radial_dist_fnc_quaternary} belong to a different species. We find that even when we treat the heavy mass particles as a different species, the structure remains the same as that of the regular KA model system.

\section{CALCULATION OF LOCAL MEAN-FIELD CAGING POTENTIAL}
\label{mean-field caging potential appendix}

 To perform the microscopic investigation of local caging potential at each frame (Eq.\ref{binary_phir} and Eq.\ref{quaternary_phir}), we have to determine the partial rdfs at a single particle level. This is done by using a sum of Gaussian to express the single particle partial rdf in a single frame, and it is calculated as follows\cite{piggi_PRL};

 \begin{equation}
 g_{ij}^{\alpha}(r) = \frac{1}{4\pi\rho r^2} \sum_{\beta} \frac{1}{\sqrt{2\pi\delta^{2}}} \exp^{-\frac{(r-r_{\alpha\beta})^2}{2\delta^2}} 
 \label{particle_rdf_eq}
 \end{equation}
\noindent
where $\rho$ is the system density, ``$\alpha$” is the particle index. The Gaussian distribution's variance ($\delta = 0.09 \sigma_{AA}$) is employed to transform the discontinuous function into a continuous form. Single particle rdf is also used to derive the direct correlation function at the single particle level from $C_{ij}(r) = -\beta u_{ij}(r) + [g_{ij}(r)-1] - \ln[g_{ij}(r)]$.

In the calculation of the caging potential, we need to obtain the product of the rdf and the direct correlation function, which leads to the calculation of the product of rdf and interaction potential. As demonstrated in a previous study \cite{mohit_paper1}, the particle level rdf produced by the Gaussian approximation has finite values at distances less than the average rdf. Thus, this range has a significant unphysical contribution at small r because of the finite value of the rdf and its product with the interaction potential, which diverges at small r. We use an approximation expression of the direct correlation function, $C_{ij}^{approx} (r)=[g_{ij}(r)-1]$, to get over this issue. In this case, the interaction potential is assumed to be equal to the potential of mean force $-\beta u_{ij}(r) = \ln(g_{ij}(r))$. It was previously shown that the theoretical prediction of structure-dynamics correlation is significantly improved by the use of $C_{ij}^{apporx}(r)$, instead of using $C_{ij}(r)$ \cite{palak_polydisperse_softness,mohit_paper2}. In this work, we always use $C_{ij}^{apporx}(r)$ in the calculation of the depth of the mean-field caging potential. The structural order parameter (SOP) is the inverse of the depth of caging potential.

The local caging potential for the ``A" particles in the ternary system can be written 
\begin{equation}
 \beta\Phi_{r}^{T}(A,\Delta r = 0) = -4\pi\rho \int r^2dr \sum_{j=1 }^3 \chi_{j}^{'} C_{1j}(r) g_{1j}(r) 
 \label{ternary_phir_A}
\end{equation}

Using a similar argument as given in the main text, the local caging potential for the ``A" particles in the modified ternary system where the ``C" particles are assumed to be pinned can be written as,

\begin{equation}
\begin{split}
 \beta \Phi_{r}^ {MT} (A,\Delta r = 0) & = -4\pi\rho \int r^2dr \Big [ \sum_{j=1}^2 \chi_{j}^{'} C_{1j}(r) g_{1j}(r) \\
 & \hspace{0.8cm} + 2 \times \sum_{j=3} \chi_{j}^{'} C_{1j}(r)g_{1j}(r) \Big ]
\end{split}
 \label{modified_ternary_phir_A}
\end{equation}

\section{ONSET TEMPERATURES OF THE MASS SYSTEMS FROM INHERENT STRUCTURE ENERGY}
\label{inherent structure}
\begin{figure}[h!]
 \centering
 \includegraphics[width=0.7\linewidth]{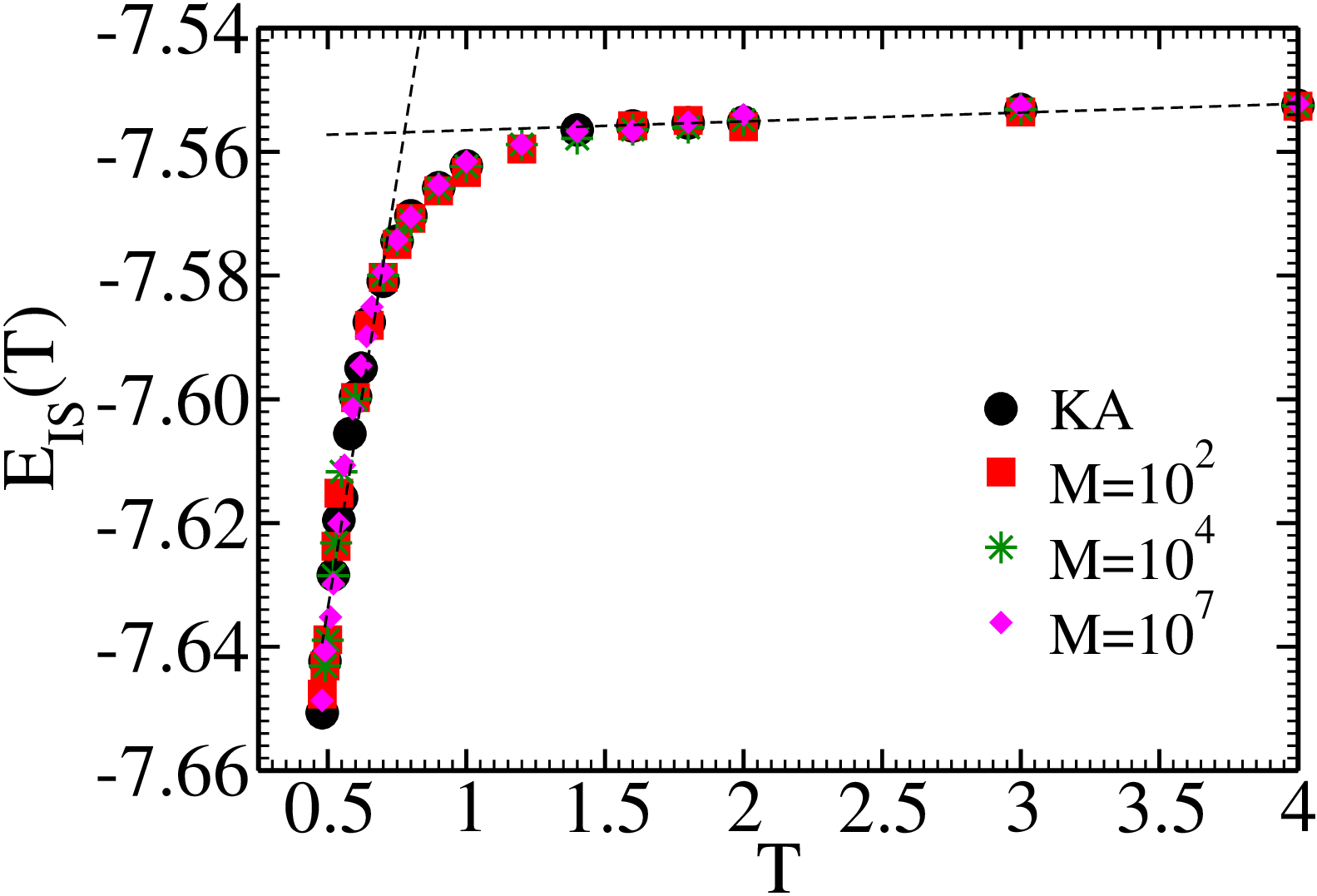}
 \caption{Inherent structure energy, $E_{IS}$ as a function of temperature (T). The onset temperature, $T_{onset}$, is the temperature where $E_{IS}$ starts to drop from its high-temperature value. The onset temperature remains the same ($T_{onset}=0.80$) for all the heavy mass systems (KA(black, circle), M = $10^2$ (red, square), M = $10^4$ (green, star), M = $10^7$ (magenta, diamond)). Here, the dotted lines are straight line fits in the high T and low T ranges.}
 \label{inherent}
\end{figure}

To estimate the onset temperature, $T_{onset}$ of the system, we use the inherent energy analysis approach \cite{sastry_inherent_1998}. In Fig. \ref{inherent}, we plot the inherent structure energy, $E_{IS}$, as a function of T. 
 We obtain the inherent structure (IS) using the fast inertial relaxation engine (FIRE) algorithm\cite{FIRE_method_for_inherent_str}
We find that the $E_{IS}$ is independent of the mass of the heavy particles. From this analysis, we can observe that the onset temperature ($T_{onset}=0.80$) remains the same with increasing masses.

{\bf ACKNOWLEDGMENT}\\
 S.~M.~B. thanks, Science and Engineering Research Board (SERB, Grant No. SPF/2021/000112 ) for the funding. The authors would like to thank Srikanth Sastry, Ganapathy Ayappan, Chandan Dasgupta, Manoj Kumar Nandi, and Ujjwal Kumar Nandi for the discussions. E.~A. Thanks, SERB, for the fellowships. M.~S. Thanks, CSIR, for the fellowships. \\[4mm]

{\bf AVAILABILITY OF DATA}\\
The data that supports the findings of this study is available from the corresponding author upon reasonable request.\\[3mm]

\textbf{References}


\begin{thebibliography}{10}

\bibitem{Theorectical_perspective_berthier_2011}
L.~Berthier and G.~Biroli,
\newblock Rev. Mod. Phys. {\bf 83}, 587 (2011).

\bibitem{stillinger}
P.~G. Debenedetti and F.~H. Stillinger,
\newblock Nature {\bf 410}, 259 (2001).

\bibitem{ANGELL}
C.~A. Angell,
\newblock Science {\bf 267}, 1924 (1995).

\bibitem{sastry_nature_1998}
F.~H.~S. Srikanth~Sastry, Pablo G.~Debenedetti,
\newblock Nature {\bf 393}, 554 (1998).

\bibitem{Debenedetti_book}
D.~P. G.,
\newblock {\em Metastable Liquids. Concepts and Principles} (Princeton Univ. Press, Princeton, 1996).

\bibitem{andrea_supercooled_liq_2009}
A.~Cavagna,
\newblock Phys. Rep. {\bf 476}, 51 (2009).

\bibitem{Kauzmann}
W.~Kauzmann,
\newblock Chem. Rev. {\bf 43}, 219 (1948).

\bibitem{kirk_woly1}
T.~R. Kirkpatrick and P.~G. Wolynes,
\newblock Phys. Rev. A {\bf 35}, 3072 (1987).

\bibitem{kirk_woly2}
T.~R. Kirkpatrick and D.~Thirumalai,
\newblock Phys. Rev. Lett. {\bf 58}, 2091 (1987).

\bibitem{kirk_thirumalai}
T.~R. Kirkpatrick and D.~Thirumalai,
\newblock Phys. Rev. B {\bf 36}, 5388 (1987).

\bibitem{krick_thirumalai_wolynes}
T.~R. Kirkpatrick, D.~Thirumalai, and P.~G. Wolynes,
\newblock Phys. Rev. A {\bf 40}, 1045 (1989).

\bibitem{xia_woly_pnas}
X.~Xia and P.~G. Wolynes,
\newblock Proc. Natl. Acad. Sci. U.S.A. {\bf 97}, 2990 (2000).

\bibitem{walter_original_pinning}
M.~Ozawa, W.~Kob, A.~Ikeda, and K.~Miyazaki,
\newblock Proc. Natl. Acad. Sci. U.S.A. {\bf 112}, 6914 (2015).

\bibitem{smarajit_chandan_dasgupta_original_pinning}
S.~Chakrabarty, S.~Karmakar, and C.~Dasgupta,
\newblock Sci. Rep. {\bf 5}, 12577 (2015).

\bibitem{Biroli_phase_diagram}
C.~Cammarota and G.~Biroli,
\newblock Proc. Natl. Acad. Sci. U.S.A. {\bf 109}, 8850 (2012).

\bibitem{Cytoskeletal_Pinning_2015}
S.~Arumugam, E.~Petrov, and P.~Schwille,
\newblock Biophys. J. {\bf 108}, 1104 (2015).

\bibitem{effect_of_random_pinning}
U.~A. Dattani, S.~Karmakar, and P.~Chaudhuri,
\newblock J. Chem. Phys. {\bf 159}, 204501 (2023).

\bibitem{palak_ujjwal_JCP}
U.~K. Nandi {\em et~al.},
\newblock J. Chem. Phys. {\bf 156}, 014503 (2022).

\bibitem{parisi_jamming_pinned_system}
C.~Brito, G.~Parisi, and F.~Zamponi,
\newblock Soft Matter {\bf 9}, 8540 (2013).

\bibitem{walter_anh}
M.~Ozawa, A.~Ikeda, K.~Miyazaki, and W.~Kob,
\newblock Phys. Rev. Lett. {\bf 121}, 205501 (2018).

\bibitem{paddy}
I.~Williams {\em et~al.},
\newblock J. Phys. Condens. Matter {\bf 30}, 094003 (2018).

\bibitem{reply_by_chandan_dasgupata}
S.~Chakrabarty, S.~Karmakar, and C.~Dasgupta,
\newblock Proc. Natl. Acad. Sci. U.S.A. {\bf 112}, E4819 (2015).

\bibitem{reply_by_kob}
M.~Ozawa, W.~Kob, A.~Ikeda, and K.~Miyazaki,
\newblock Proc. Natl. Acad. Sci. U.S.A. {\bf 112}, E4821 (2015).

\bibitem{smarajit_soft_pinning_pnas_2024}
R.~Das, B.~P. Bhowmik, A.~B. Puthirath, T.~N. Narayanan, and S.~Karmakar,
\newblock PNAS Nexus {\bf 2}, pgad277 (2023).

\bibitem{copling_plasma_2017}
T.~Fujimoto and I.~Parmryd,
\newblock Front. Cell Dev. Biol. {\bf 4}, 155 (2017).

\bibitem{soft_pinning_of_liquid_2015}
L.~Feriani, L.~Cristofolini, and P.~Cicuta,
\newblock Chemistry and Physics of Lipids {\bf 185}, 78 (2015).

\bibitem{interleaflet_coupling_2011}
G.~G. Putzel, M.~J. Uline, I.~Szleifer, and M.~Schick,
\newblock Biophys. J {\bf 100}, 996 (2011).

\bibitem{saurish}
S.~Chakrabarty and R.~Ni,
\newblock J. Chem. Phys. {\bf 152}, 234502 (2020).

\bibitem{sayantan_fickian_2017}
S.~Acharya, U.~K. Nandi, and S.~Maitra~Bhattacharyya,
\newblock J. Chem. Phys. {\bf 146}, 134504 (2017).

\bibitem{G_ayappa_2020}
V.~Varadarajan, R.~Desikan, and K.~G. Ayappa,
\newblock Soft Matter {\bf 16}, 4840 (2020).

\bibitem{G_ayappa_2022}
S.~Choudhury, V.~Ananthanarayanan, and K.~G. Ayappa,
\newblock Soft Matter {\bf 18}, 4483 (2022).

\bibitem{Kob_Andersen_1995}
W.~Kob and H.~C. Andersen,
\newblock Phys. Rev. E {\bf 51}, 4626 (1995).

\bibitem{palak_JCP_2024}
P.~Patel and S.~Maitra~Bhattacharyya,
\newblock J. Chem. Phys. {\bf 160}, 164501 (2024).

\bibitem{manoj_PRL_2021}
M.~K. Nandi and S.~M. Bhattacharyya,
\newblock Phys. Rev. Lett. {\bf 126}, 208001 (2021).

\bibitem{mohit_paper1}
M.~Sharma, M.~K. Nandi, and S.~M. Bhattacharyya,
\newblock Phys. Rev. E {\bf 105}, 044604 (2022).

\bibitem{kob-andersen}
W.~Kob and H.~C. Andersen,
\newblock Phys. Rev. E {\bf 51}, 4626 (1995).

\bibitem{lammps}
D.~L. {Majure} {\em et~al.},
\newblock Large-Scale Atomic/Molecular Massively Parallel Simulator (LAMMPS) Simulations of the Effects of Chirality and Diameter on the Pullout Force in a Carbon Nanotube Bundle , IEEE , 201 (2008).

\bibitem{thesis_palak}
P.~Patel,
\newblock Correlation between structure, entropy and dynamics in multi-species systems, Thesis  (2023).

\bibitem{walter_Tg}
W.~Kob,
\newblock Course 5: Supercooled liquids, the glass transition, and computer simulations,
\newblock in {\em Slow Relaxations and nonequilibrium dynamics in condensed matter}, edited by J.-L. Barrat, M.~Feigelman, J.~Kurchan, and J.~Dalibard, pp. 199--269, Berlin, Heidelberg, 2003, Springer Berlin Heidelberg.

\bibitem{glotzer_2003}
N.~Lačević, F.~W. Starr, T.~B. Schrøder, and S.~C. Glotzer,
\newblock J. Chem. Phys. {\bf 119}, 7372 (2003).

\bibitem{siladitya_2011}
S.~Sengupta, F.~Vasconcelos, F.~Affouard, and S.~Sastry,
\newblock J. Chem. Phys. {\bf 135}, 194503 (2011).

\bibitem{VFT}
L.~S. Garca-Coln, L.~F. del Castillo, and P.~Goldstein,
\newblock Phys. Rev. B {\bf 40}, 7040 (1989).

\bibitem{Vogel}
H.~Vogel,
\newblock Phys. Z. {\bf 22} (1921).

\bibitem{Tammann}
G.~Tammann and W.~Hesse,
\newblock Z. Anorg. Allg. Chem. {\bf 156}, 245 (1926).

\bibitem{Fulcher}
G.~S. Fulcher,
\newblock J. Am. Ceram. Soc. {\bf 8}, 339 (1925).

\bibitem{Harowell}
A.~W. Cooper, H.~Perry, P.~Harrowell, and D.~R. Reichman,
\newblock Nat. Phys. {\bf 4}, 711 (2008).

\bibitem{Liu15}
E.~D. Cubuk {\em et~al.},
\newblock Phys. Rev. Lett. {\bf 114}, 108001 (2015).

\bibitem{Liu16}
S.~S. Schoenholz, E.~D. Cubuk, D.~M. Sussman, E.~Kaxiras, and A.~J. Liu,
\newblock Nat. Phys. {\bf 12}, 469 (2016).

\bibitem{Liu17}
E.~D. Cubuk {\em et~al.},
\newblock Science {\bf 358}, 1033 (2017).

\bibitem{Yodh19}
X.~Ma {\em et~al.},
\newblock Phys. Rev. Lett. {\bf 122}, 028001 (2019).

\bibitem{Kohli20}
V.~Bapst {\em et~al.},
\newblock Nat. Phys. {\bf 16}, 448 (2020).

\bibitem{Tanaka18}
T.~S. Ingebrigtsen and H.~Tanaka,
\newblock Proc. Natl. Acad. Sci. U.S.A. {\bf 115}, 87 (2018).

\bibitem{Tanaka18-1}
H.~Tong and H.~Tanaka,
\newblock Phys. Rev. X {\bf 8}, 011041 (2018).

\bibitem{Tanaka20}
H.~Tong and H.~Tanaka,
\newblock Phys. Rev. Lett. {\bf 124}, 225501 (2020).

\bibitem{mohit_paper2}
M.~Sharma, M.~K. Nandi, and S.~Maitra~Bhattacharyya,
\newblock J. Chem. Phys. {\bf 159}, 104502 (2023).

\bibitem{palak_polydisperse_softness}
P.~Patel, M.~Sharma, and S.~Maitra~Bhattacharyya,
\newblock J. Chem. Phys. {\bf 159}, 044501 (2023).

\bibitem{Ramakrishnan_Yussouff_1979}
T.~V. Ramakrishnan and M.~Yussouff,
\newblock Phys. Rev. B {\bf 19}, 2775 (1979).

\bibitem{Hansen_and_McDonald}
J.~P. Hansen and I.~R. McDonald,
\newblock The Theory of Simple Liquids, 2nd ed. ,Academic, London  (1986).

\bibitem{Cooper_IC_2004}
A.~Widmer-Cooper, P.~Harrowell, and H.~Fynewever,
\newblock Phys. Rev. Lett. {\bf 93}, 135701 (2004).

\bibitem{Cooper_IC_2006}
A.~Widmer-Cooper and P.~Harrowell,
\newblock Phys. Rev. Lett. {\bf 96}, 185701 (2006).

\bibitem{Bertheir_IC}
L.~Berthier and R.~L. Jack,
\newblock Phys. Rev. E {\bf 76}, 041509 (2007).

\bibitem{piggi_PRL}
P.~M. Piaggi, O.~Valsson, and M.~Parrinello,
\newblock Phys. Rev. Lett. {\bf 119}, 015701 (2017).

\bibitem{sastry_inherent_1998}
S.~Sastry, P.~G. Debenedetti, and F.~H. Stillinger,
\newblock Nature {\bf 393}, 554 (1998).

\bibitem{FIRE_method_for_inherent_str}
J.~Guénolé {\em et~al.},
\newblock Comput. Mater. Sci {\bf 175}, 109584 (2020).

\end{thebibliography}

\end{document}